\pdfoutput=1
\documentclass[letterpaper]{article}
\usepackage[totalwidth=480pt, totalheight=680pt]{geometry}

\usepackage[utf8]{inputenc} 
\usepackage[T1]{fontenc}    
\usepackage[dvipsnames]{xcolor}         
\usepackage[pagebackref,colorlinks=true,urlcolor=blue,linkcolor=blue,citecolor=OliveGreen]{hyperref}               
\usepackage{url}            
\usepackage{booktabs}       
\usepackage{amsfonts}       
\usepackage{nicefrac}       
\usepackage{microtype}      
\usepackage[T1]{fontenc}
\usepackage{lmodern}
\title{Generalizations of Length Limited Huffman Coding for Hierarchical Memory Settings}

\author{
Shashwat Banchhor\\
Dept. of Comp. Science, Indian Institute of Technology, New Delhi, India\\
\texttt{shashwatbanchhor12@gmail.com}
\and
Rishikesh Gajjala\\
Indian Institute of Science, Bangalore, India\\
Dept. of Comp. Science, Indian Institute of Technology, New Delhi, India\\
\texttt{rishikeshg@iisc.ac.in}
\and
Yogish Sabharwal\\
IBM Research, New Delhi, India\\
\texttt{ysabharwal@in.ibm.com}
\and
Sandeep Sen\footnote{Currently on leave from Dept. of Comp. Science, Indian Institute of Technology, New Delhi, India}\\
Dept. of Comp. Science, Shiv Nadar University, U.P, India\\
\texttt{ssen@snu.edu.in}
}
\date{}

\usepackage{amsthm}
\usepackage{multirow}
\usepackage{color}
\usepackage{lineno}
\usepackage{makeidx}
\usepackage{xspace}
\usepackage[title]{appendix}
\usepackage{forest}
\usepackage{epsfig}
\usepackage{amsmath}
\usepackage{amssymb}
\usepackage{url}

\usepackage{graphicx}
\usepackage{float}

\usepackage[linesnumbered,ruled,vlined]{algorithm2e}
\usepackage{tikz}
\usetikzlibrary{calc}

\usepackage{xcolor}
\usepackage{todonotes}

\DeclareMathOperator*{\argmin}{arg\,min}

\usepackage{caption}

\newtheorem{Definition}{Definition}

\definecolor{darkred}{rgb}{1, 0.1, 0.3}
\definecolor{darkblue}{rgb}{0.1, 0.1, 1}
\definecolor{darkgreen}{rgb}{0,0.6,0.5}

\newcommand {\mm}[1] {\ifmmode{#1}\else{\mbox{\(#1\)}}\fi}

\newcommand{\htparam}     {{\sc D}}

\newcommand{\alphabet}  {{C}}

\newcommand{\fd}[1]     {F_{{#1}}}

\usepackage{amssymb,amsthm,amsmath}
\usepackage{bbm}
\usepackage{graphicx}
\usepackage{enumerate}
\usepackage{subfigure}
\usepackage[bottom]{footmisc}
\usepackage{comment} 

\usepackage[capitalise,nameinlink]{cleveref}
\usepackage{thmtools}
\usepackage{thm-restate}
\theoremstyle{plain}
\newtheorem{theorem}{Theorem}[section]
\newtheorem{lemma}[theorem]{Lemma}
\newtheorem{proposition}[theorem]{Proposition}

\newtheorem{corollary}[theorem]{Corollary}

\theoremstyle{remark}

\usepackage{tikz}
\usetikzlibrary{calc, graphs, graphs.standard, shapes, arrows, positioning, decorations.pathreplacing, decorations.markings, decorations.pathmorphing, fit, matrix, patterns, shapes.misc}

\begin{document}

\maketitle

\newcommand{\dT}        {{d_T}}
\newcommand{\dTp}        {{d_{T'}}}
\newcommand{\dTopt}     {{d_{T^*}}}
\newcommand{\len}       {{len}}
\newcommand{\freq}      {{freq}}
\newcommand{\cost}      {{P}}
\newcommand{\DOPT}      {{\sc COPT}}
\newcommand{\SLLHC}     {{\sc Soft-LLHC}}
\newcommand{\bdT}       {{W}}
\newcommand{\PDOPT}     {{\sc Par-DOPT}}
\newcommand{\LLHF}      {{\sc LLHF}}
\newcommand{\LLHC}      {{\sc LLHC}}
\newcommand{\GSLLHC}    {{\sc Gen-LLHC}}
\newcommand{\DPTOPT}    {\sc PDec}
\newcommand{\DPLLHF}    {\sc LDec}
\newcommand{\DPF}       {\sc F}
\newcommand{\DPLLHFLEN} {\sc LLen}
\newcommand{\eat}[1]    {}
\newcommand{\struct}[1] {{\sc struct(#1)}}
\newcommand{\lvl}       {\ell}
\newcommand{\dual}      {\Tilde{D}}
\newcommand{\pnlty}     {p}
\newcommand{\codelength} {c}
\newcommand{\objT}      {F}
\newcommand{\objc}      {f}
\newcommand{\prefx}      {S}

\begin{abstract}
In this paper, we study the problem of designing prefix-free encoding schemes
having minimum average code length that can be decoded
efficiently under a decode cost model that captures memory hierarchy induced cost functions.
We also study a special case of this problem that is closely related to the 
length limited Huffman coding (LLHC) problem; we call
this the {\em soft-length limited Huffman coding}  problem. In this version, 
there is a penalty associated with each of the $n$ characters of the alphabet
whose encodings exceed a specified bound $D$($\leq n$)
where the penalty increases linearly with the length of the encoding 
beyond $D$. 
The goal of the problem is to find a prefix-free encoding
having minimum average code length and total penalty within a
pre-specified bound ${\cal P}$.
This generalizes the LLHC problem.
We present an algorithm to solve this problem that runs in time 
$O( nD )$.
We study a further generalization in which the penalty 
function and the
objective function can both be arbitrary monotonically non-decreasing functions of the codeword length.
We provide dynamic programming based exact and PTAS algorithms for this setting.
\end{abstract}

\section{Introduction}
Data compression algorithms aim to reduce the number of bits required to represent data
in order to save storage capacity, speed up file transfer, and decrease costs for storage 
hardware and network bandwidth.
Compression techniques are primarily divided into two categories: lossless and lossy.
Lossless compression enables data to be restored to its original state, without the loss of a single bit of data,
when it is uncompressed (decoded).
Huffman encoding is a basic and popular approach for lossless data compression 
based on variable length prefix-free encoding \cite{huffman},
where the characters of the alphabet are encoded with variable length codewords and no character encoding is a
prefix of another.
Huffman encoding is widely used in many applications including file compression (e.g. GZIP \cite{pkzip}, PKZIP \cite{pkzip}, BZIP2 \cite{bzip2}, etc.) and image and video storage formats (JPEG \cite{jpeg}, PNG \cite{png}, MP3 \cite{mp3}, etc.).

Traversal of a Huffman tree to decode compressed data has an inherent cost proportional to the path length that can be prohibitively slow for many real time applications.
One such application is inference task in deep learning.
As the sizes of deep learning models are quite large, 
smaller models are obtained by using Huffman coding
in conjunction with other techniques to reduce the memory consumption \cite{HanMD15}.
The model is decoded in real-time when inference has to be performed.
In such settings, 
it is acceptable to trade-off the compression ratio for improved decode time 
as this is a critical aspect for a good user experience.
Since the data is encoded only once, it may be beneficial to spend the extra
time in suitably encoding data to expedite decoding.

To avoid repeated sequential path traversals of the Huffman tree, we can exploit the indirect addressing capabilities of the RAM model by using lookup tables;
code trees are employed where small tables are used to represent subtrees \cite{634683}.  
If $w$ bits are used (called the width of the table),
then the size of the table is $2^w$. So we partition the code into prefixes of 
smaller lengths when the tree is not balanced, to economize space.
If a prefix of the lookup bits forms a valid code word, then the table entry points to the corresponding
code word and the input slides ahead by the number of bits used in the encoding of the code word.
Otherwise, the table entry points to another table where a lookup is performed with the next
fixed number of bits (possibly different than $w$) of the input; this is repeated until a valid word is decoded.
This is illustrated in Figure~\ref{fig:multi-tables}.

 This scheme is further complicated by the memory hierarchy that limits the 
storage at the faster levels of memory and has increasing latencies as we 
access deeper tables.
The prefix tree can be viewed as multiple levels of blocks where each block corresponds to a lookup table
used during decoding. 
Figure~\ref{fig:blocking} demonstrates the concept of blocking 
where we assume that the blocks that require the same number of indirections 
have similar latencies.
This problem can be formulated as follows:

\begin{figure}[t!]
\centering
\begin{minipage}{.6\textwidth}
  \includegraphics[width=3.3in]{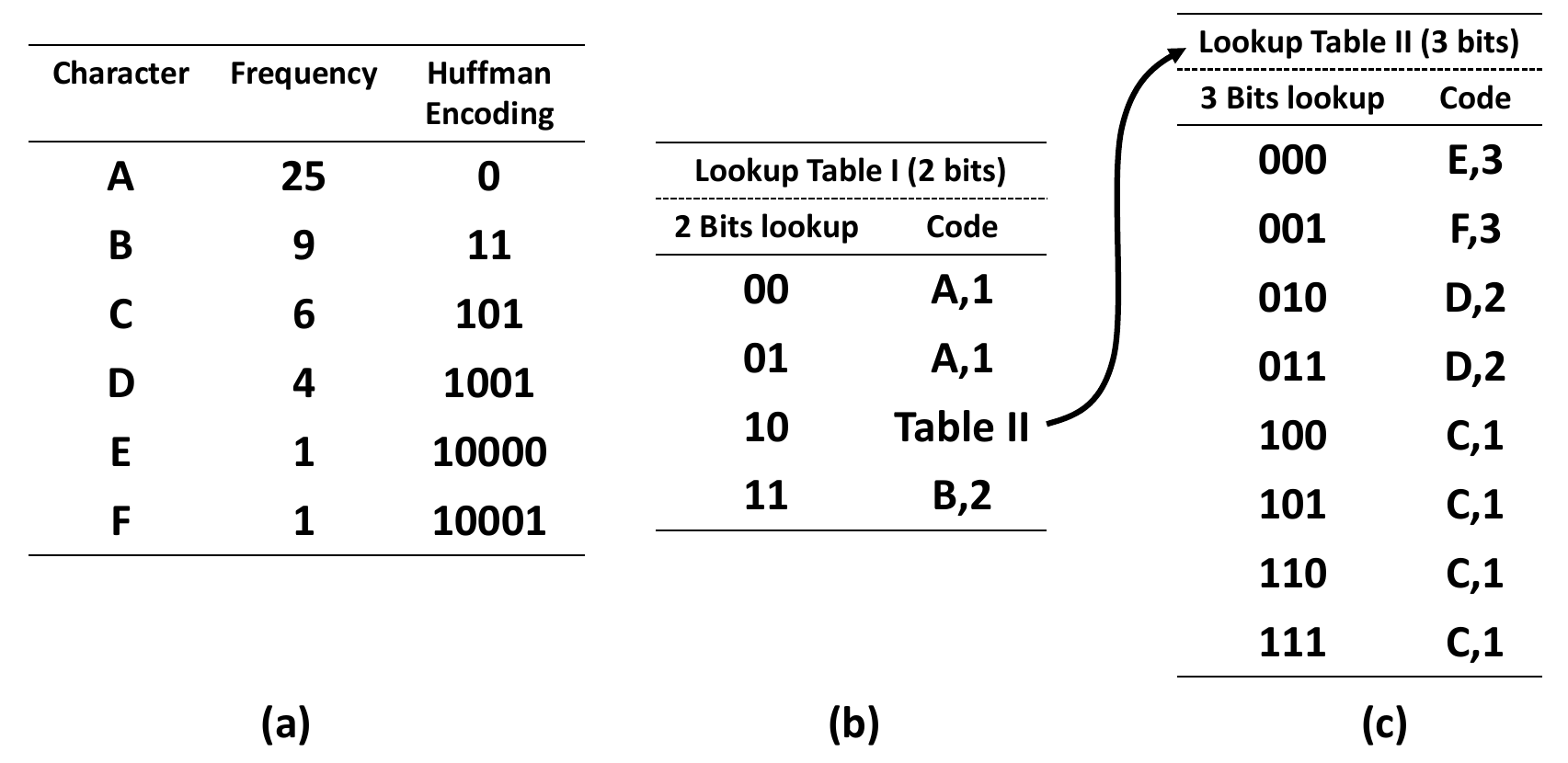}
  \captionof{figure}{\small Consider an alphabet with frequencies and Huffman encoding as shown in Table (a).
    Tables (b) and (c) illustrate the $1^{st}$ and $2^{nd}$ level lookup tables of width 2 and 3 bits respectively. 
    }
    \label{fig:multi-tables}
\end{minipage}%
\ \ \ \ 
\begin{minipage}{.35\textwidth}
    \centering
  \includegraphics[width=2in]{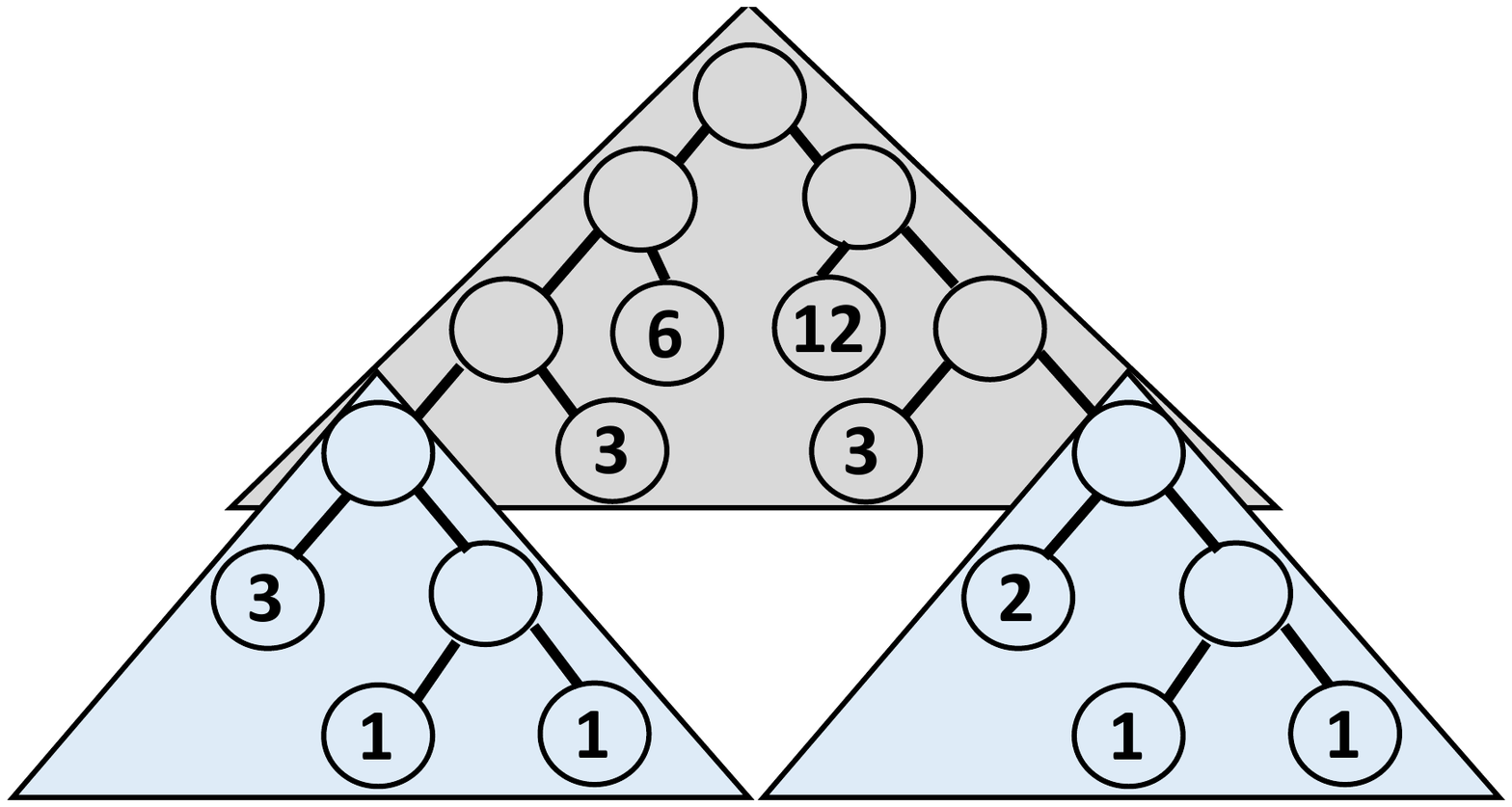}
    \ \\
    \captionof{figure}{Illustration of blocking scheme: $<$$(3,1),(2,1)$$>$.}
    \label{fig:blocking}
    \ \\
\end{minipage}
\end{figure}
Consider an alphabet $\alphabet$ such that the size of alphabet, $|\alphabet|=n$. 
For each character $c$ in $\alphabet$, let the attribute $\freq(c)$ denote the frequency of $c$ 
in the input data to be encoded. 
Given a prefix tree $T$ corresponding to a prefix-free code for $\alphabet$,
let $\dT(c)$ denote the depth of the leaf corresponding to the encoding of $c$ in the tree. 
Note that $\dT(c)$ is also the length of the codeword for character $c$. 
The code length of the encoding represented by the tree $T$ is given by
\begin{eqnarray}
\label{def:len}
\len(T)=\sum_{c \in \alphabet} \freq(c) \cdot \dT(c)
\end{eqnarray}
Define a \textit{blocking scheme} of $m$ \textit{block levels} as a sequence of $m$ block parameters,
 $<$$(w_1,q_1),$$(w_2,q_2),$
 $\ldots,$$(w_m,q_m)$$>$,
where $w_j$ and $q_j$ specify the width and the access cost of a block, 
respectively, at \textit{block level} $j$ in the tree.
For a blocking scheme, 
the number of memory hierarchies
is the number of times the  access cost changes when traversing the blocks in order.
For a character $c$ having depth $\dT(c)$ in a prefix tree $T$,
the cost of looking up (decoding) the character under the scheme $BS$, $\delta_T(c)$,
is given by the total sum of the cost of accessing the blocks starting from the 
first \textit{block level} up to the \textit{block level} to which the character belongs, i.e.,
$\delta_T(c) \ = \sum_{i \le \bdT(c)} q_i$ where 
$\bdT(c) \ = \ \argmin_h \left\{ \sum_{j=1}^{h} w_j \ge \dT(c) \right\}$.
The total decode time of the encoding for a prefix tree $T$ is given by:
$\delta(T)\ =\ \sum_{c \in \alphabet} \freq(c)\cdot \delta_T(c)$.

\begin{quote}
\par\noindent
{\bf Problem Definition ({\DOPT}): }
Given a blocking scheme $BS$ and parameter ${\Delta}$, called the {\em permitted cost},
the goal of our problem is to determine a prefix tree, $T$,
that minimizes the code length, $len(T)$, subject to 
$\delta(T) \le {\Delta}$. 
We call this the {\em code optimal prefix tree problem} and denote it by {\DOPT}(${\Delta}$).
With slight abuse of notation, 
we shall also refer to the code length of the associated solution as {\DOPT}(${\Delta}$).
\end{quote}

We present an exact and a PTAS algorithm for the {$\DOPT(\Delta)$} problem:
\begin{theorem}
\label{thm:memhei}
\begin{quote}
\begin{enumerate}
\item[{\em \bf (a)}] There exists a dynamic programming based algorithm to solve the $\DOPT(\Delta)$ problem that runs
in time $O(n^{2+m})$ for $m$ block levels.

\label{thm:constanthei} 
\item[{\em \bf (b)}] For the case where the number of block levels, $m$, is a constant,
there exists an algorithm that returns a prefix tree having  code-length $\le (1+\epsilon) \DOPT({\Delta})$. 
The running time of the algorithm
is $O\left(\dfrac{n^2}{\epsilon}\max\left(\dfrac{1}{\epsilon^2}, {\log^2(n)}\right) \right)$. \\

\end{enumerate}
\end{quote}
\end{theorem}

Another technique for optimizing the decode time that is popular in practice
was proposed by Moffat and Turpin \cite{634683}.
Their algorithm looks up one entry of an {\em offset} array (sequentially) for every bit of the compressed data
read from the input. 
To speed up their algorithm, they use a lookup table using a fixed number
of bits from the input.
This is then followed by looking up an entry of the offset array for every subsequent bit of the input.
The lookup table is often kept in fast memory as compared to the offset array.
The overall decode time can be optimized by accommodating more words in the lookup table.
This can be modeled as a special case of the {\DOPT} problem where the memory hierarchy comprises of only two levels. 
The first level corresponds to a cache or scratchpad having constant memory access cost.
The second level corresponds to the main memory for which every access incurs a cost of $q$.
This corresponds to a blocking scheme of $\langle$$(w_1,z),$$(w_2,q),$$(w_2,q),$$\ldots$ $\rangle$.
Any entry of the prefix tree residing in the cache can be accessed with constant cost $z$ and 
thereafter every entry in the main memory is accessed with cost $q$.
Intuitively, if the codes cannot fit into the topmost block, 
we need a design that will minimize the number of higher level (deep) blocks. 
Having a hard-bound on the code word length has been previously dealt under Length Limited Huffman Code (LLHC) problem\cite{karp}; 
we define a variation to deal with the current problem using a notion of penalties.

LLHC is a well studied variant of Huffman coding motivated by the construction of 
optimal prefix-free codes under certain practical conditions \cite{garey}
such as computer file searching and text retrieval systems \cite{ZobelM95}.
The LLHC($C,D$) problem outputs a prefix-free encoding over alphabet $\alphabet$, whose lengths are bounded by $D$
such that the code length is minimized.
The encoding length bound, $D$, is a hard bound in the LLHC problem and is naturally bounded by the size of the alphabet $n$.
Consider a soft version of the LLHC problem, where there is a penalty associated
with the character encodings exceeding bound $D$ that increases linearly with the length of the
encoding. Given a bound on the admissible penalty, the goal of the problem is to find a prefix-free encoding
having minimum code length and penalty within the specified admissible bound.
We note that this problem also allows us to consider settings where the desired character encoding length $D$ 
is smaller than $\log n$; this is impossible in the {\LLHC} setting because of the information theoretic bottleneck.

We next define this generalized version of the LLHC problem more formally.
For a character having depth $\lambda$ in a prefix tree $T$,
we associate a penalty, $\pnlty(.)$ as follows:
$$\pnlty(\lambda) =
\left\{
	\begin{array}{ll}
		z  & \mbox{if } \lambda \leq D \\
		z + q \cdot (\lambda-D) & \mbox{if } \lambda > D
	\end{array}
\right.
$$
for some constants $z$ and $q$. Here, $z$ is a constant cost for character encodings having length no more than $D$
and $q$ is the penalty for every extra encoding bit used beyond $D$. The reader may note that this is a simplification from the natural blocking model where the number of bits may be more than 1. However, this assumption allows us to exploit certain properties leading to very fast solutions that are likely to work well in practice.
The penalty of the prefix tree is the sum of the penalties of all 
the characters weighted by their frequencies, i.e., 
\begin{eqnarray}
\label{def:costtree}
\cost(T)=\sum_{c \in \alphabet} \freq(c) \cdot \pnlty(\dT(c)).
\end{eqnarray}
\begin{quote}
\par\noindent
{\bf Problem Definition ({\SLLHC}): }
Given parameters $z$, $q$ \& $D$, which define the penalty function $\pnlty(.)$
and a penalty bound ${\cal P}$,
the goal of the {\em Soft length limited Huffman coding problem}, denoted {\SLLHC}(${\cal P},z,q,D$),
is to determine a prefix tree, $T$, that minimizes the code length $\len(T)$ subject to 
$\cost(T) \le {\cal P}$. 
\end{quote}

Figure~\ref{fig:example} illustrates the Huffman coding for an alphabet ${\alphabet}$,
the corresponding LLHC and {\SLLHC} when $D=3$. 
We note that LLHC is a special case of this problem wherein $z=0$, $q=1$ and ${\cal P}=0$.
This setting does not allow for any penalty, and constrains character encodings to have length $\le D$.
Thus {\SLLHC} is a generalization of the LLHC problem. We present a fast algorithm for the {\SLLHC} problem:

\begin{figure}[t!]
    \centering
    \begin{minipage}{0.28\textwidth}
        \centering
        \resizebox{1.0\textwidth}{!}{
        \begin{tikzpicture}[
        level 1/.style={level distance=8mm,sibling distance=20mm},
        level 2/.style={level distance=8mm,sibling distance=12mm},
        level 3/.style={level distance=8mm,sibling distance=12mm},
        font=\scriptsize,inner sep=2pt,every node/.style={draw,circle,minimum size=3ex}]
        noda/.style={draw,circle,minimum size=3ex}]
        \node {67} 
        child {node[rectangle] {34} 
            edge from parent }
        child   {node {33} 
                    child {node[rectangle] {17} 
                         edge from parent }
                    child {node {16} 
                        child{ node[rectangle] {11}
                            edge from parent}
                        child {node{5}
                            child{ node[rectangle] {3}
                                edge from parent}
                            child {node{2}
                                child{ node[rectangle]{1}
                                    edge from parent}
                                child{ node[rectangle]{1}
                                    edge from parent}
                                    }
                            }
                    }
                } ;
        \end{tikzpicture}}\\
        {\small (a)}
    \end{minipage}
    \begin{minipage}{0.4\textwidth}
        \centering
        \resizebox{1.0\textwidth}{!}{
        \begin{tikzpicture}[
        level 1/.style={level distance=10mm,sibling distance=36mm},
        level 2/.style={level distance=10mm,sibling distance=24mm},
        level 3/.style={level distance=10mm,sibling distance=16mm},
        font=\scriptsize,inner sep=2pt,every node/.style={draw,circle,minimum size=3ex}]
        
        \node {67} 
        child {node {51} 
                child {node[rectangle]{34}} 
                child {node[rectangle]{17}
                }
            edge from parent }
        child   {node {16} 
                    child {node {14} 
                        child {node[rectangle] {11}} 
                        child {node[rectangle]{3}}
                    }
                    child {node {2}
                        child {node[rectangle] {1}}
                        child {node[rectangle] {1}}}
                } ;
    \end{tikzpicture}}\\ \ \\ \ \\
    {\small (b)}

    \end{minipage}
    \begin{minipage}{0.24\textwidth}
        \centering
        \resizebox{1.0\textwidth}{!}{
        \begin{tikzpicture}[
        level 1/.style={level distance=8mm,sibling distance=20mm},
        level 2/.style={level distance=8mm,sibling distance=12mm},
        level 3/.style={level distance=8mm,sibling distance=6mm},
        font=\scriptsize,inner sep=2pt,every node/.style={draw,circle,minimum size=3ex}]
        noda/.style={draw,circle,minimum size=3ex}]
        \node {67} 
        child {node[rectangle] {34} 
            edge from parent }
        child   {node {33} 
                    child {node {28} 
                        child{
                        node[rectangle]{17}
                                    edge from parent}
                                child{ node[rectangle]{11}
                                    edge from parent}
                         edge from parent }
                    child {node {5} 
                        child{ node[rectangle] {3}
                            edge from parent}
                        child {node{2}
                            child{ node[rectangle] {1}
                                edge from parent}
                            child{ node[rectangle] {1}
                                edge from parent}
                            }
                    }
                } ;
        \end{tikzpicture}}\\
        {\small (c)}    
\end{minipage}

    \caption{Consider an alphabet with $6$ characters with frequencies ${1,1,3,11,17,34}$ and $\htparam = 3$. Any character with depth $w \le 3$ bits has a penalty of $z$ unit whereas characters with depth $w>3$ bits have penalty $z + q \cdot (w-D)$. 
    (a) illustrates the corresponding Huffman tree that has code length of $5\cdot1+5\cdot1+4\cdot3+3\cdot11+2\cdot17+1\cdot34 = 123$ and penalty of $(z+2q)\cdot1+(z+2q)\cdot1+(z+q)\cdot3+z\cdot11+z\cdot17+z\cdot34 = 67z+7q$.
    (b) Illustrates the LLHC prefix tree with higher code length of 
    $3\cdot1+3\cdot1+3\cdot3+3\cdot11+2\cdot34+2\cdot17=150$
    but a penalty of $z\cdot1+z\cdot1+z\cdot3+z\cdot11+z\cdot17+z\cdot34=67z$. 
    (c) Illustrates the soft-LLHC prefix tree with code length of
    $4\cdot1+4\cdot1+3\cdot3+3\cdot11+2\cdot17+1\cdot34=128$ and penalty of $(z+q)\cdot1+(z+q)\cdot1+z\cdot3+z\cdot11+z\cdot17+z\cdot34=67z+2q$.
    }
    \label{fig:example}
\end{figure}
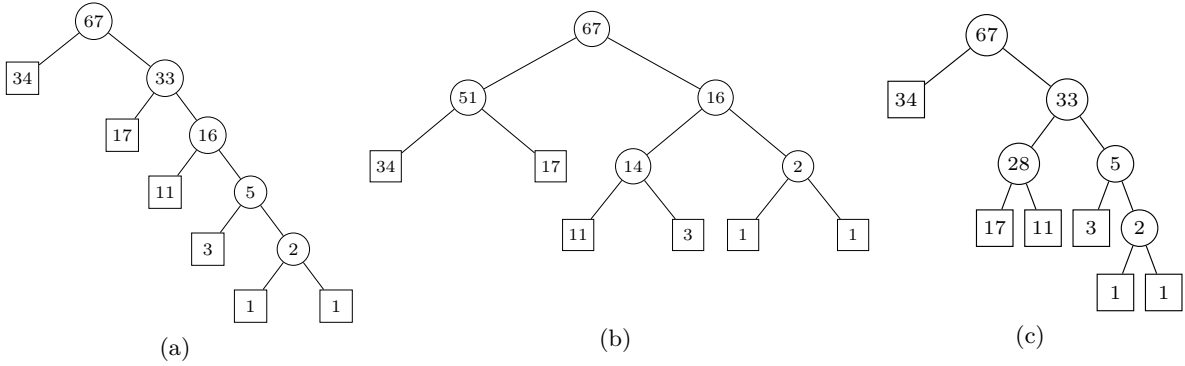

\begin{theorem}
\label{softllhc}
There exists an algorithm to solve the {\SLLHC}(${\cal P},z,q,D$) problem with running time $O( nD )$
when the characters of $C$ are given in sorted order of frequencies. For the case when $D=o(\log n)$, the running time of the algorithm can be bounded by $O( n + D2^{D} )$.
\end{theorem}

Note that a special case of our {\DOPT} problem with two levels of memory hierarchy for $BS=\langle$ $(w_1,z),(1,q),(1,q)\cdots$ $\rangle$
can be mapped to the {\SLLHC} problem by taking $D=w_1$ and $\cal{P}=$ ${\Delta}$.

Lastly, we study a more generalized version of the {\SLLHC} problem that also generalizes the {\DOPT} problem.
In this problem, the penalty and
cost functions can be any monotonically non-decreasing function of the code length.

We next define this problem formally.

\par\noindent
\begin{quote}
{\bf Problem Definition ({\GSLLHC}): }
Given parameters ${\cal P}$, called the {\em penalty bound}, a penalty function $\pnlty(\cdot)$ and an
objective function $f(\cdot)$ that are both monotonically non-decreasing functions,
the goal of the {\em Generalized length limited Huffman coding problem}, 
denoted {\GSLLHC}(${\cal P}, \pnlty(\cdot), f(\cdot)$),
is to determine a prefix tree, $T$, that minimizes 
$$F(T)=\sum_{c \in \alphabet} \freq(c) \cdot f(\dT(c)).$$
subject to the penalty being bounded by the specified penalty bound, i.e.,
$$\cost(T)=\sum_{c \in \alphabet} \freq(c) \cdot \pnlty(\dT(c)) \le {\cal P}.$$
\end{quote}
Note that in {\GSLLHC} the penalty function is not necessarily linear, as it was in {\SLLHC}. 

Also note that the {\DOPT} problem can be modeled as the {\GSLLHC} problem by taking the penalty function as
$\pnlty(\dT(c)) = \delta_T(c)$, ${\cal P}$ as ${\Delta}$ and the function $f$ mapping 
to the code length, i.e., $f(\dT(c)) = \dT(c)$. {Note that the effect of $BS$ is handled in the way $\pnlty(\dT(c))$ is defined}. We present the following results for the {\GSLLHC} problem:

\begin{theorem}
\label{gllhc}

\label{thm:exact} 
\begin{quote}
\begin{enumerate}  
\item[{\em \bf (a)}] There exists a dynamic programming algorithm to solve 
the 
{\GSLLHC}$({\cal P}, \pnlty(\cdot), f(\cdot))$ problem that runs in $O(n^3 \cdot {\cal P})$ time.
\label{thm:ptas} 
\item[{\em \bf (b)}] There exists a dynamic programming algorithm that returns a prefix-tree having 
objective value at most $(1+\epsilon)$ times that of the optimal solution to {\GSLLHC}$({\cal P}, \pnlty(\cdot), f(\cdot))$
and penalty no more than ${\cal P}$. 
The running time of this algorithm is $O(n^4 / \epsilon)$.
\end{enumerate}
\end{quote}
\end{theorem}

{\em Remark 1.} Note that while the running time in Theorem \ref{softllhc} has no 
dependence on ${\cal P}$, Theorem \ref{thm:exact}(a)  is not a 
strictly polynomial time algorithm for super polynomial values of ${\cal P}$. 

{\em Remark 2.} Theorem \ref{thm:exact}(a),(b) assume the functions $\pnlty(\cdot), f(\cdot)$ can be computed in $O(1)$ time.

{\em Hardness.} {Note that it follows from Theorem \ref{softllhc} that the {\SLLHC} problem is in $P$ 
as $D$ can be at most $n$. We do not have a hardness result for the {\GSLLHC} problem though we present a PTAS for the problem in Theorem \ref{thm:ptas}}. { The {\DOPT} problem is a special case of the {\GSLLHC} problem for which we give an algorithm which runs in polynomial time when the number of block levels is constant.}

\subsection{Related Work}
\label{related}
The first algorithm for LLHC was due to Karp\cite{karp} and was based on an integer linear programming
formulation. Gilbert\cite{gilbert} then gave an enumeration based algorithm for {\LLHC}. Both these algorithms
had exponential running time. Later Hu and Tan\cite{hutan} gave an $O(nD2^D)$ time Dynamic Programming algorithm. 
Note that $D$ is bounded by $n$ in the worst case.
In 1974, Garey\cite{garey} presented the first polynomial time algorithm, 
running in time $O(n^2D)$ for the case of binary encoded alphabets.
Larmore\cite{larmore} combined techniques of \cite{hutan} and \cite{garey} to give an algorithm 
with running time $O(n^{3/2}D \log^{1/2}n)$ for the binary case.
Larmore and Hirschberg \cite{llhf} then designed a completely new algorithm with running time $O(nD)$; 
this algorithm was based on a reduction to the coin collector's problem, which was then solved using
a technique they called the Package-Merge algorithm.
There have been several subsequent works that have improved the running time further for the special case
when $D=\omega(\log n)$ to $O(n\sqrt{D\log{n}}+n\log{n})$ by Aggarwal, Schieber and Tokuyama \cite{aggarwal1994finding} and to $n2^{O(\sqrt{\log{D}\log\log{n}})}$ by Schieber \cite{schieber1998computing}.
Baer \cite{baer2005source} studied a variant of the Huffman coding problem
wherein there is a continuous (strictly) monotonic increasing cost (penalty) function,
called Campbell penalties \cite{campbell1966definition}, associated with the length of a character encoding; 
the goal of the problem is to minimize the ``mean'' length
of the cost function over all the characters of the alphabet.
We note that this problem seeks to minimize an objective different from the code length,
thereby addressing a different setting compared to Huffman coding, {\LLHC} and our {\SLLHC} problems.
In particular, the {\SLLHC} problem seeks to minimize the code length constrained by a budget on the 
admissible penalty.

Generalized cost functions for building Huffman trees have been studied before.
Fujiwara and Jacobs \cite{FujiwaraJ14} studied the Generalized Huffman Tree (GHT) problem in which the 
cost of each encoded character depends on its depth in the tree by an arbitrary function. 
Here the goal is to determine a prefix tree, $T$, that minimizes 
$\sum_{i = 1}^{|\alphabet|} f_{i}(\dT(c_{i}))$ for the GHT problem and minimizes $\max_{i = 1}^{|\alphabet|} f_{i}(\dT(c_{i}))$ for the Max-GHT problem.
This is a further generalization of our cost function, where a separate function is associated with each character.

On the other hand, the {\SLLHC} problem corresponds to optimizing the objective
function allowing deviations from the individual code lengths 
for which we provide bi-criterion results that are 
novel to the best of our knowledge.
We do note however that the {\LLHC} problem is a special case of both the {\SLLHC} problem 
(as specified earlier) as well as the GHT problem 
(by taking the cost function to be $\infty$ when depth exceeds $D$ and equal to frequency times depth otherwise).

Fujiwara and Jacobs \cite{FujiwaraJ14} further prove that the Max-GHT problem is 
NP-hard when the cost functions are allowed to be arbitrary {and provide a polynomial time algorithm when the cost functions are non-decreasing}. 
We observe that the hardness result crucially depends on the the prefix tree being a complete binary tree.
However, we show that for certain functions, the optimal prefix tree for Max-GHT need not necessarily be
a complete binary tree (see Section ~\ref{appendix_fujiwara}).
As a matter of fact, we present a simple polynomial time construction for the relaxed version of 
Max-GHT by reducing the Max-GHT problem with arbitrary functions into Max-GHT 
problem with non-decreasing functions in $O(n^2)$ time. 
Using the polynomial time algorithm of Fujiwara and Jacobs \cite{FujiwaraJ14} for the case when the cost functions are non-decreasing, this actually yields a polynomial time algorithm for the case of arbitrary functions as well.
This result is captured in the following theorem.

\begin{theorem}
\label{thm:fujiwara}
There is an $O(n^2\log{n})$ algorithm for Max-GHT with arbitrary functions.
\end{theorem}

{\bf Organization of the paper.} 
Our algorithms build on the dynamic program for Huffman codes {proposed by Larmore and Przytycka\cite{huff_rec_prelim} and extended in Golin\cite{golin}}. 
This algorithm is discussed in Section~\ref{sec:prelims}.
In Section~\ref{monge_sllhc_algo}, we first present our algorithm for the simplest of the problems, {\SLLHC};
this provides the proof for Theorem~\ref{softllhc}.
In Section~\ref{sec:genresults}, we discuss the algorithmic approach for the generalized version of the {\GSLLHC} problem; this corresponds to Theorem~\ref{thm:ptas}.
In Section~\ref{copt_algorithms}, we present the algorithms for the {\DOPT} problem.
This is presented last as the proofs reuse results from the algorithm for {\GSLLHC}. In section~\ref{appendix_fujiwara}, we present the proof for Theorem \ref{thm:fujiwara}.
We end with concluding remarks in Section~\ref{sec:concl}.
\section{Preliminaries: a DP for Huffman codes}
\label{sec:prelims}

\begin{figure}[t!]
    \centering
    \begin{minipage}{0.6\textwidth}
        \centering
        \resizebox{1.0\textwidth}{!}{
    \includegraphics[width=10cm]{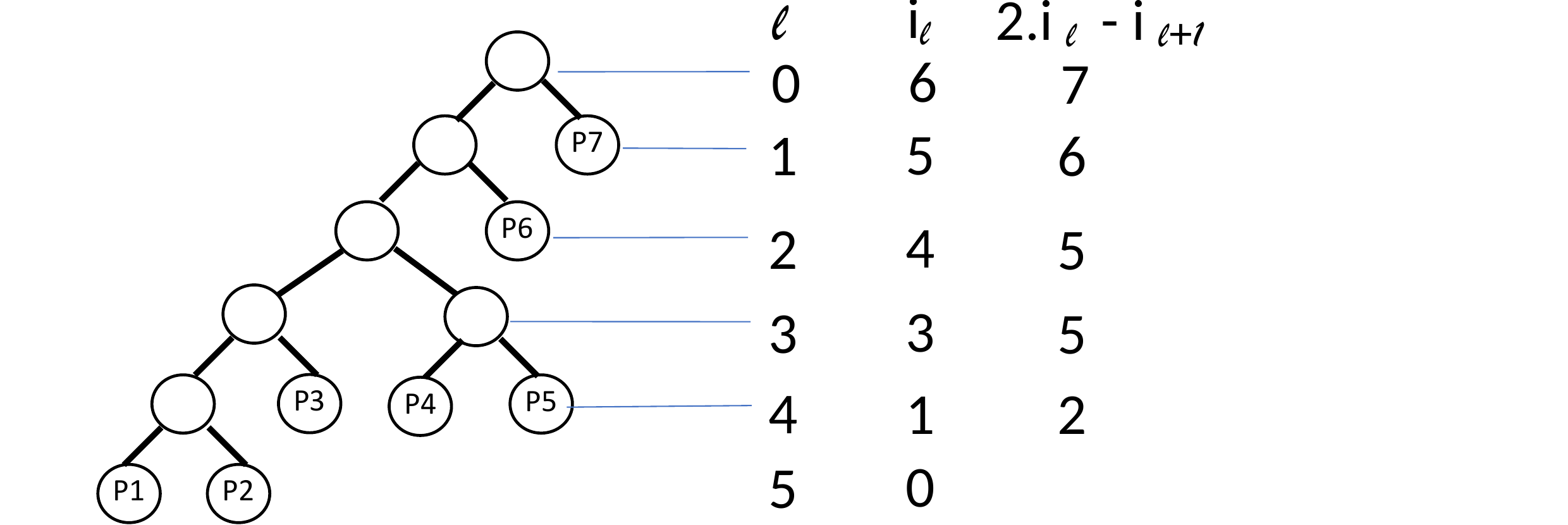}}
    \caption{Illustration of the calculation of the number of characters below level $\lvl$ ($2i_\lvl - i_{\lvl+1}$). 
    This figure is taken from \cite{golin}.}
    \label{fig:golin_example}
    \end{minipage} \ \ \ \ \ \ \ \ 
    \begin{minipage}{0.33\textwidth}%
    \centering
    \ \\
    \ \\
   \resizebox{0.80\textwidth}{!}{
    \includegraphics[width=10cm,height=10cm,keepaspectratio,trim={5cm 2cm 0 0},clip]{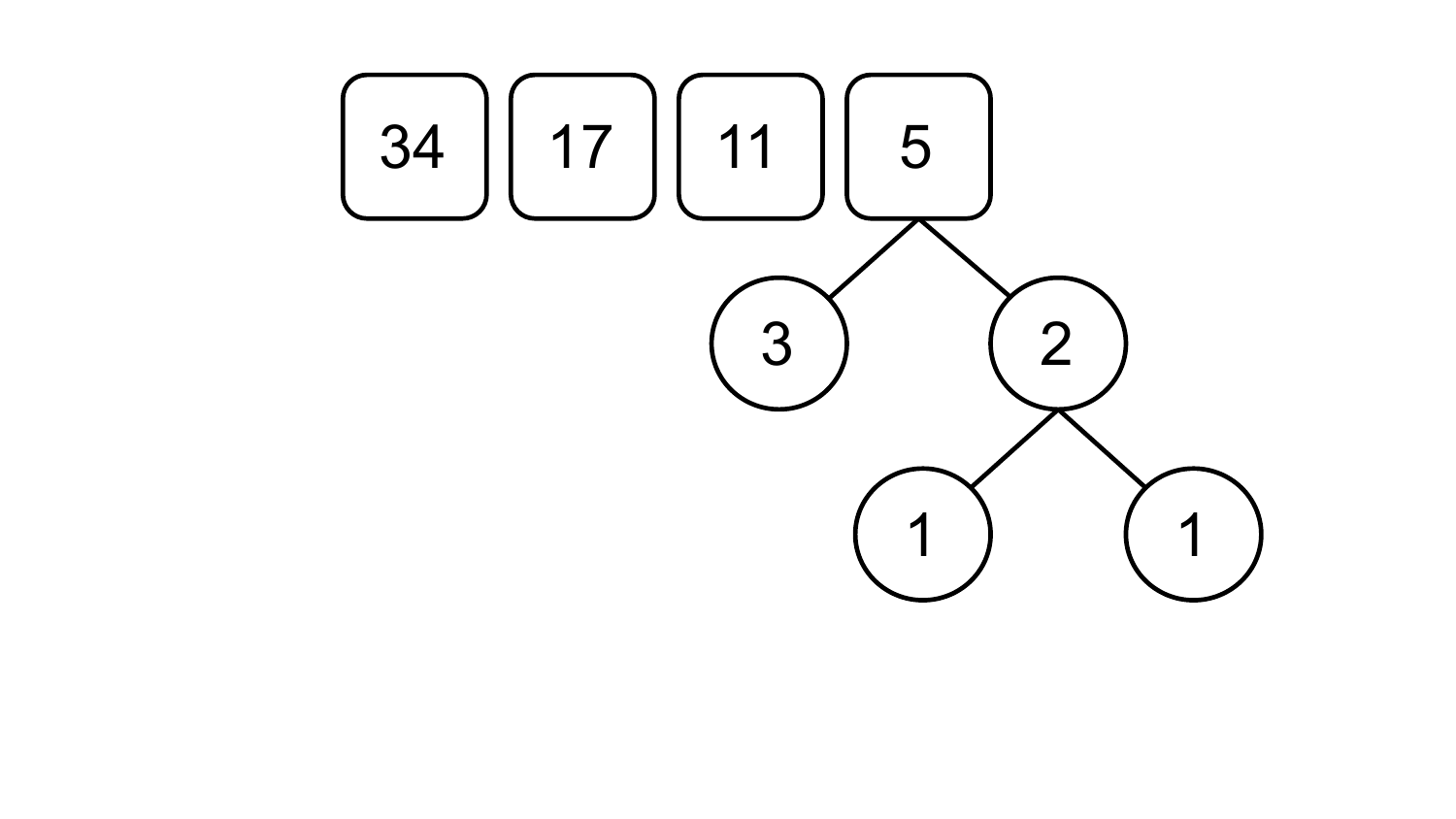}
    }
    \caption{The {\bf ${\mathbf 3}$-level forest} to the tree $T$, shown in Figure~\ref{fig:example}(a)}
    \label{fig:example2}
    \end{minipage}
\end{figure}

Consider a prefix tree, $T$. The nodes of $T$ can be classified as either leaf nodes (i.e.,
nodes with no child nodes), or internal nodes (i.e., nodes with exactly $2$ child nodes).
Leaf nodes represent characters of the alphabet.
Let $\dT(u)$ denote the depth of any node in the tree, $T$ (with the root being at depth $0$). 
The depth (or height) of the tree, denoted $h(T)$ is the maximum depth of any node in the tree, i.e.,
$h(T) = \max_{u \in T} \{ \dT(u) \}$.
We use the variable $\lvl$ to refer to the level starting from the top of the tree ($\lvl=0$ for the root).
Further, let $i_\lvl$ denote the number of internal nodes at or deeper than level $\lvl$. This is illustrated in Figure~\ref{fig:golin_example}. 
The following proposition relates the number of characters below some level with the number of internal nodes
at different levels.
\begin{proposition}
\label{charbelow}
The number of characters below (deeper than) level $\lvl$ is $2i_\lvl - i_{\lvl+1}$.
\end{proposition}
The formal proof of the proposition can be found in \cite{golin}.
The intuitive idea is as follows:
to form each internal node we need two child nodes (can be either internal or leaf). 
Hence, for $i_\lvl$ internal nodes we would require $2 i_\lvl$ nodes at or deeper than level $(\lvl+1)$.
Since of these $2 i_\lvl$ nodes $i_{\lvl+1}$ are internal nodes at or deeper than level $(\lvl+1)$, 
the number of characters or leaf nodes below level $\lvl$ must be $2 i_\lvl - i_{\lvl+1}$.

The following Theorem is an adaptation of a result from Golin and Zhang\cite{golin} that 
specifies a condition for us to be able to 
construct a valid prefix tree. {Note that Golin and Zhang \cite{golin} did not require the condition that 
$\forall \lvl \le h-2, n \geq (2 i_\lvl - i_{\lvl+1}) \geq (2 i_{\lvl+1} - i_{\lvl+2})$.
They instead proved that any sequence that is an optimal solution to the LLHC problem 
corresponds to a valid prefix tree (Lemma 2 and 8 in \cite{golin}). 
We instead show that this extra condition is necessary and sufficient, 
for any $\cal{I}$ to correspond to a valid full binary prefix tree.}

\eat
{We prove a more generalised version of Theorem \ref{valid}
}
\begin{theorem}
\label{valid_forest}
Given a decreasing sequence of integers, 
$\cal{I}=\langle$$i_k,i_{k+1},\ldots,i_h=0$$\rangle \ \ $, such that $ \forall \lvl \le h-2, \ \ n \geq (2 i_\lvl - i_{\lvl+1}) \geq (2 i_{\lvl+1} - i_{\lvl+2}) $ and $i_k \leq n-1$
we can construct a forest, rooted at level $k$, such that the number of internal nodes
at or below level $\lvl$ is $i_\lvl$.
\end{theorem}

\begin{proof}
We prove this constructively by induction. For the sequence $\cal{I'}=\langle$$i_{h-1},i_h=0$$\rangle$, we can construct a forest with $i_{h-1}$ trees, each containing one internal node and two leaves. Since this forest has no internal nodes at or below level $h$, we have $i_{h}=0$ . Also, since the only internal nodes are the roots of the trees at level $h-1$, we have $i_{h-1}$ internal nodes at or below level $h-1$. Further, as $2i_{h-1} \leq n$ we have sufficient characters to construct this forest). \\
Now, let us assume there is a valid forest corresponding to the sequence $\cal{I'}=\langle$$i_{k+1},\cdots i_h=0$$\rangle$. Note that this forest has $i_{k+1}-i_{k+2} > 0$ trees. We now add another $(2i_{k}-i_{k+1})-(2i_{k+1}-i_{k+2})$ leaves (characters) at level $k+1$ and construct a forest with $i_{k} - i_{k-1}$ trees, having a total of $i_{k}$ internal nodes. Note that $(2i_{k}-i_{k+1})-(2i_{k+1}-i_{k+2}) \geq 0$ and $(2i_{k}-i_{k+1}) \leq n$, hence, we have sufficient characters to create such a forest. This proves the theorem.
\end{proof}

Corollary \ref{valid} follows from Theorem \ref{valid_forest} when $k=0$ and $i_0=n-1$.
\begin{corollary}

\label{valid}
\label{valid_sequence_proof}
Given a decreasing sequence of integers, 
$\cal{I}=\langle$$i_0=n-1,i_1,\ldots,i_h=0$$\rangle$, such that $\ \forall \lvl \le h-2: \ n \geq (2 i_\lvl - i_{\lvl+1}) \geq (2 i_{\lvl+1} - i_{\lvl+2})$,
we can construct a prefix tree
of height $h$ such that the number of internal nodes
at or below level $\lvl$ is $i_\lvl$.
\end{corollary}

Observe that in any optimal prefix-tree, a character with higher frequency cannot appear lower 
than a character having lower frequency (otherwise we could swap them leading to an improved codelength).
Using this fact,
the following result from Golin and Zhang\cite{golin} helps us to rewrite the 
code length of the code represented by a prefix tree as the sum of contributions of prefix sums at each level.
\begin{theorem}\label{prefixtree}
Let $\prefx=[\prefx_1,\prefx_2\cdots \prefx_n]$ be prefix sum array of frequencies, where $\prefx_i=\sum_{j=1}^{i}freq(j)$ and frequencies are sorted in increasing order
of the depths of the characters in the tree $T$. 
Then the code length of the tree, $T$, can be written as a sum of $h$ prefix sums, 
where each sum represents the code length contribution by each level of the tree, i.e.,
$ len(T)= \sum_{\lvl=0}^{h-1}\prefx_{2 i_\lvl - i_{\lvl+1}}$.
\end{theorem}
The formal proof can be found in \cite{golin}. The intuitive idea is as follows:

by the definitions of $len(T)$ and $\dT(c)$ (Eqn~\ref{def:len} and depth of character $c$ in tree $T$), we have
$$ len(T)=\sum_{c \in \alphabet} \freq(c) \cdot \dT(c) = \sum_{c \in \alphabet} \sum_{\lvl=1}^{\dT(c)} \freq(c)$$
By rearranging the summation over each level and using Proposition \ref{charbelow}, we get
$ len(T)=\sum_{\lvl=0}^{h-1} \sum_{j=1}^{2 i_\lvl - i_{\lvl+1}}freq(j)$
and viewing the inner sum as prefix sum, we get
$ len(T) = \sum_{\lvl=0}^{h-1}\prefx_{2 i_\lvl - i_{\lvl+1}}$.

The goal of Golin and Zhang\cite{golin} is to determine a prefix tree, $T$, for which the code length, i.e., $len(T)$ is minimum.
The idea of their dynamic program is as follows.
Let $H(i)$ denote the minimum code length amongst all forests having exactly $i$ internal nodes.  
Then $H(n-1)$ yields the optimal code length.
As mentioned in Corollary \ref{valid}, it suffices to obtain a sequence of $i_\lvl$'s to determine the prefix tree.
Suppose that $i_\lvl=i$ and $i_{\lvl+1}=j$, then due to Larmore and Przytycka \cite{huff_rec_prelim} we have, 
$H(i) = H(j) + \prefx_{2 i-j}$. 
This allows us to determine the optimal values of $i_\lvl$'s as follows.
We initialize $H$ to $\infty$ for all entries and then use the following recurrence:
$$H(i) = \min_{\substack{ j \in [max(0,2i-n),i-1] \\ \& \ 2i-j \ \ge \ 2j-k}} H(j) + \prefx_{2 i-j}$$
where $k$ is the recursive index used in populating $H(j)$,
i.e., $H(j)$ was minimized for $H(k) + \prefx_{2j-k}$
(this can be recorded in a separate table);
the condition {($2i-j \ \ge \ 2j-k$) ensures that the number of leaves below the root level in the structure with $i$ internal nodes is greater than or equal to that in the structure with $j$ internal nodes.}
Here, $H(0)$ and $\prefx_0$ are initialized to $0$.

{\em Time complexity:}
As there are $n$ entries of $H$ and each entry requires $O(n)$ computations to compare the recurrences, 
the algorithm takes $O(n^2)$ time. Using the concavity of $S_i$, this was improved to $O(n)$ time in \cite{parallelhuffman} by filling the cells using Concave Least weight Subsequence (CLWS), { which can be solved using SMAWK algorithm as a subroutine in $O(n)$ time \cite{CLWS}}. 

{\em Remark:} We use a slightly different notion of level than \cite{golin}. 
While \cite{golin} considers levels starting with the bottom most level as $0$ and increasing up to the root, 
we consider levels to start with $0$ from the root and increasing down the tree.
The above theorems and lemmas have been rephrased accordingly.

\section{Algorithm for the Soft-LLHC (\SLLHC) Problem}
\label{monge_sllhc_algo}

Note that {\SLLHC}(${\cal P},z,q,D$) can be reformulated as {\SLLHC}(${\cal P'},0,1,D$) 
by taking ${\cal P'}=\frac{1}{q} \cdot ({{\cal P}-z \sum_{c \in \alphabet} freq(c)})$. 
Here on we work with this reformulation of the problem.

Consider the structure of the prefix tree, $T$, in a solution to the {\SLLHC} problem.
Recall that a character with higher frequency cannot appear below a character with lower frequency.
\eat{First we note that as in the case of Huffman trees, in any optimal prefix-tree for {\SLLHC}, 
a character with higher frequency cannot appear lower 
than a character having lower frequency (otherwise we could swap them leading to an improved solution).}
We can view the tree as comprising of levels starting with level 0 at the root.
We define the {\bf ${\mathbf d}$-level forest of ${\mathbf T}$, denoted $\fd{d}(T)$}, 
to be the forest induced on $T$, obtained by removing all the internal nodes 
having depth less than $d$ (along with their incident edges). Note that  
the leaf nodes having depth less than or equal to $d$ become singleton trees in $\fd{d}(T)$.
See Figure~\ref{fig:example2} for an illustration.
Let $d_{\fd{d}(T)}(c)$ denote depth of character $c$ in this forest.
Note that in the reformulated version of our problem, 
the penalty of the entire tree $T$ is equal to the codelength of the forest rooted at level $D$ for any tree. 
Hence for $d \le D$, the penalty of the tree $T$ can be written as
$$\sum_{c \in \alphabet: d_{\fd{d}(T)}(c) > D-d } \left( d_{\fd{d}(T)}(c) - (D-d) \right) \cdot freq(c).$$

We maintain a table $\overline{H}$ of size ${\cal C} \times D$.
Intuitively, for $0 \le i < |{\cal C}|$ and $1 \le d \le D$, 
an entry $\overline{H}(i,d)$ of this table tries to capture the structure of the $d$-level forest, $\fd{d}(T')$, 
corresponding to the best prefix tree, $T'$, for which $\fd{d}(T')$ comprises $i$ internal nodes.
More precisely, an entry $\overline{H}(i,d)$ of this table represents the minimum amongst the code lengths
of all forests (over the alphabet ${\cal C}$) comprising of exactly $i$ internal nodes and 
additionally satisfying the condition that the penalty condition is not violated, i.e., 
$$\sum_{c \in \alphabet: d_{\fd{d}(T)}(c) > D-d } \left( d_{\fd{d}(T)}(c) - (D-d) \right) \cdot freq(c)) \le {\cal P}'.$$

Note that for $d<D$, the $d$ level forest $\fd{d}(T^*)$ in the optimal tree $T^*$ is formed 
by introducing new internal nodes that combine some of the trees of the forest $\fd{d+1}(T^*)$
(by merging their roots pairwise to form new internal nodes). 
From the previous section, the code length of a prefix-tree of height $h$ can be represented as sum of $h$ prefix-sums. This is applicable for the {\SLLHC} problem as well.
Thus, the values of the table $\overline{H}$ can be computed as follows.
Initialize all entries of $\overline{H}$ to $\infty$ and then
use the following recurrence for $d < D$:
$$\overline{H}(i,d) = \min_{\substack{ j \in [max(0,2i-n),i-1] \\ \& \ 2i-j \ \ge \ 2j-k}} \overline{H}(j,d+1) + \prefx_{2i-j}.$$
Here $k$ corresponds to the recursive index used in populating $\overline{H}(j,d+1)$,
i.e., $\overline{H}(j,d+1)$ was minimized for $\overline{H}(k,d+2) + \prefx_{2j-k}$
(this can be recorded in a separate table).

Note that at level $D$, an entry $\overline{H}(i,D)$ corresponds to the minimum code length amongst all 
forests having exactly $i$ internal nodes and penalty no more than ${\cal P}'$. Since
the penalty of this tree corresponds exactly to its codelength
(as $z=0$ and $q=1$ for the reformulated {\SLLHC}), this actually corresponds
exactly to the entry $H(i)$, provided $H(i) \le {\cal P}'$
and we can thus initialize $\overline{H}(i,D)=H(i)$.
Note that if $H(i) > {\cal P}$, then there does not exist a tree with penalty less than ${\cal P}$ 
and having $i$ internal nodes with depth at least $D$;
thus we can set $\overline{H}(i,D)=\infty$ in this case. 
The final solution is then obtained from the entry $\overline{H}(n-1,0)$.
The prefix tree can be constructed by alluding to Theorem~\ref{valid}.

{\em Time complexity.} The entries of $H$ can be computed in time $O(n)$ as discussed previously.
For computing $\overline{H}$, there are $nd$ cells and each cell takes $O(n)$ time to fill using the recurrence above. 
Hence the running time is $O(n^2D)$. This time can be improved by employing properties of Monge matrices.
This is discussed next.

\par\noindent
{\bf Improving the running time using Monge property.}
Monge property is a discrete extension of concavity which allows for the speeding up of several algorithms\cite{monge}. 
SMAWK is one such classical algorithm, using which row-minima can be found. It was used by Golin and Zhang\cite{golin} to solve the LLHC problem. We follow a similar approach. Consider the recurrence
$$\widehat{H}(i,d) = \min_{j \in [max(0,2i-n),i-1]} \widehat{H}(j,d+1) + \prefx_{2i-j}.$$
Note that we drop the condition $2i-j \ge 2j-k$ from the recurrence for $\overline{H}$.
This is because we can argue that the optimal sequence will correspond to a valid prefix-tree.
As all the solutions of $\widehat{H}$ satisfy the penalty constraint, we only minimize the code length. If the optimal sequence doesn't correspond to a valid prefix-tree, it is possible to construct a sequence using Lemma 8 in \cite{golin}, with a smaller value of $\sum_{\lvl=0}^{D-1}\prefx_{2 i_\lvl - i_{\lvl+1}}$, leading to a contradiction.

Now consider a new implicit matrix $M^{(d)}$ such that for all $0 \le i,j \le n$: $M_{i,j}^{(d)}=\widehat{H}(j,d+1) + S_{2i-j}$
when $0\leq 2i-j\leq n$ and $\infty$ when $2i-j>n$ or $2i-j<0$.
We show that $M^{(d)}$ is a Monge matrix. This follows from the following Lemma.
\begin{lemma}
\label{lemma:monge}
$M_{i,j}^{(d)}+M_{i+1,j+1}^{(d)} \leq M_{i+1,j}^{(d)}+M_{i,j+1}^{(d)}$. \\
(Note that SMAWK allows for this condition to be satisfied when both sides evaluate to $\infty$).

\end{lemma}
\begin{proof}
We consider the following three (exhaustive) cases:

\par\noindent
(I) {\em When $2i-j < 1$:} $M_{i,j+1}^{(d)}$ is $\infty$ and thus the result holds by definition (as $2i-(j+1) < 0$).

\par\noindent
(II) {\em When $2i-j > n-2$:} $M_{i+1,j}^{(d)}$ is $\infty$ and thus the result holds by definition (as $2(i+1)-j >n$).

\par\noindent
(III) {\em When $1 \leq 2i-j \leq n-2$:} 
entries $M_{i,j}^{(d)},M_{i+1,j+1}^{(d)},M_{i+1,j}^{(d)},M_{i,j+1}^{(d)}$ are defined and we have:
\begin{align*}
( M_{i,j}^{(d)}+M_{i+1,j+1}^{(d)} ) \ & \ - \ ( M_{i+1,j}^{(d)} + M_{i,j+1}^{(d)} )\\
&\leq \ (\widehat{H}(j,d+1) + S_{2i-j} + \widehat{H}(j+1, d+1) + S_{2i-j+1}) \\
& \ \ \ \ \ \ \ \  - \ (\widehat{H}(j, d+1) + S_{2i-j+2} + \widehat{H}(j+1, d+1) + S_{2i-j-1}) \\
& = \ S_{2i-j} +  S_{2i-j+1} -  S_{2i-j+2} - S_{2i-j-1} \\
& = \ freq(c_{2i-j})-freq(c_{2i-j+2}) \ \ \ \leq \ \ \ 0
\end{align*}
where $c_{i}$ is the $i$th least frequent character and thus the result holds.
\end{proof}
Observe that, by definition, $\widehat{H}(i,d) = \min\limits_{0 \leq j \leq i}M_{i,j}^{(d)} = \min\limits_{0 \leq j \leq n}M_{i,j}^{(d)} $.
The Monge property on $M^{(d)}$ implies that 
the SMAWK algorithm\cite{SMAWK} can solve for the row minima of each $M^{(d)}$ matrix in $O(n)$ time. 
Thus our algorithm repeats the process of finding row minima of each $M^{(d)}$ matrix for $d=D-1$ to $0$
to obtain the minima corresponding to $\widehat{H}(.,d)$.
Thus solving for $D$ such matrices takes $O(nD)$ time(See Algorithm \ref{alg:soft_llhc_faster_algo}).
The number of internal nodes with depth at most 
$D$ is bounded by $2^{D+1}$ and hence the computation of $H(i)$ takes $O(n)$ time. 
Thus, the run time can be bounded by $O(n+D2^{D})$ when $D=o(\log n)$.
\begin{algorithm}[h!]
\DontPrintSemicolon 
\KwIn{Weighted Alphabet $\alphabet=\{c_1, c_2, \ldots, c_n\}$; Penalty bound ${\cal P}$;}
\KwOut{Minimum code length prefix tree having penalty less than ${\cal P}$}

$\prefx$ $\gets$ Prefix~sum~array~of~sorted~frequencies\\
$H(i)$ $\gets$ From CLWS for all $i \in [0,n-1]$

\For{$i \gets ~0~\textbf{to}~n-1$}{
\uIf{$H(i) ~ \leq {\cal P}$}{
                $\widehat{H}(i,D) = H(i)$
                }
\uIf{$H(i) ~ > {\cal P}$}{
                $\widehat{H}(i,D) = \infty$
                }
}
\For{$d \gets ~D-1~\textbf{to}~0$}{
$SMAWK(M^{(d)}) \text{ uses } \widehat{H}(i,d+1) \text{ and computes } \widehat{H}(i,d)$
}

$C^*$ $\gets$ $\widehat{H}(n-1,0)$\\

$T^*$ $\gets$ Obtain the prefix tree by following the parent pointers of $C^*$ \\
\Return{$T^*$};
\caption{(for \textbf{Theorem \ref{softllhc}})}
\label{alg:soft_llhc_faster_algo}
\end{algorithm}
\section{Algorithms for the Generalized LLHC (\GSLLHC) Problem}
\label{sec:genresults}

We build on the ideas of Golin and Zhang\cite{golin} (see Section~\ref{sec:prelims} for details).
By extending their construction, we show that for {\GSLLHC}(${\cal P}, 
\pnlty(\cdot), f(\cdot)$), the objective value $\objT(T)$, for any tree $T$, 
can also be written as a sum of $h$ terms. The $i^{th}$ term representing 
the product of the sum of the frequencies of all the leaf nodes with depth 
less than or equal to $i$ and the difference in the objective values at depth 
$i$ and $i-1$, that is $f(i)-f(i-1)$ ($f(0)=0$). 
However our goal is to minimize the
objective value, $\objT(T)$, of the tree. We handle the penalty bound involved 
by maintaining an extra parameter in our proposed dynamic program. We store 
structures with the minimum objective value having penalty less than the new 
parameter (corresponding to the admissible values of penalty bound) introduced. Using this formulation we obtain an exact algorithm referred to in Theorem \ref{thm:exact}(a). 

The running time of the 
exact algorithm is $O(n^3 \cdot {\cal P})$ which may be super 
polynomial in $n$ for large values of ${\cal P}$. 
Subsequently, we are able to bound 
the number of feasible penalty values using standard rounding techniques and 
get an approximate algorithm which runs in $O(n^4 / \epsilon)$ and has code length no more than $(1+\epsilon)$ time the optimal value. 
We prove a slightly generalized variant of the problem, denoted  {\GSLLHC}$^*$(${\cal P}, \pnlty(\cdot), f(\cdot), h$)
that takes an additional parameter $h$ representing a height bound and determines
a prefix tree $T$ of height at most $h$ that minimizes $F(T)$ subject to the penalty bound as before.
We show that
\begin{theorem}
\label{gllhc2}
\label{thm:exact2} 
There exists a dynamic programming algorithm that returns a prefix-tree having height at most $h$ and
objective value at most $(1+\epsilon)$ times that of the optimal solution to 
{\GSLLHC}$^*({\cal P}, \pnlty(\cdot), f(\cdot), h)$ and penalty $\le {\cal P}$ with running time of $O(n^2h^2 / \epsilon)$. 
\end{theorem}
Theorem~\ref{thm:exact}(b) follows by taking the parameter $h$ as $n$ as that is the maximum height possible.

The details of the algorithms and proofs are
presented in following subsections.

Note that unlike the {\GSLLHC} problem, the 
{\SLLHC} has strictly polynomial running time as we 
use ${\cal P}$ only to filter and remove the infeasible solutions.

{As mentioned before, we do not have a hardness result for the {\GSLLHC} problem. 
We note that proving hardness is challenging for several problems related to Huffman coding. 
For instance, hardness results are not known for Huffman coding with unequal letter costs\cite{GolinKY02} that admit a PTAS.
As another instance, we have shown that the hardness result for a closely related problem, MAX-GHT, 
due to Fujiwara and Jacobs\cite{FujiwaraJ14} in prior literature is not correct (See Theorem~\ref{thm:fujiwara} and 
the associated discussion in Section~\ref{related}).
}

\eat{
\section{{Proof of Theorem \ref{valid}}}
\label{valid_sequence_proof}
We prove a more generalised version of Theorem \ref{valid}
\begin{theorem}
\label{valid_forest}
Given a decreasing sequence of integers, 
$\cal{I}=<$$i_k,i_{k+1},\ldots,i_h=0$$> \ \ $, such that $ \forall \lvl \le h-2, \ \ n \geq (2 i_\lvl - i_{\lvl+1}) \geq (2 i_{\lvl+1} - i_{\lvl+2}) $ and $i_k \leq n-1$
we can construct a forest, rooted at level $k$, such that the number of internal nodes
at or below level $\lvl$ is $i_\lvl$.
\end{theorem}
\begin{proof}
We prove this constructively by induction. For the sequence $\cal{I'}=<$$i_{h-1},i_h=0$$ > \ \ $, we can construct a forest with $i_{h-1}$ trees, each containing one internal node and two leaves. Since this forest has no internal nodes at or below level $h$, we have $i_{h}=0$ and since the only internal nodes are the roots of the trees at level $h-1$, we have $i_{h-1}$ internal nodes at or below level $h-1$. Further, as $2i_{h-1} \leq n$ we have sufficient characters to construct this forest). \\
Now, let us assume there is a valid forest corresponding to the sequence $\cal{I'}=<$$i_{k+1},\ldots,i_h=0$$> \ \ $. Note that this forest has $i_{k+1}-i_{k+2} > 0$ trees. We now add another $(2i_{k}-i_{k+1})-(2i_{k+1}-i_{k+2})$ leaves(characters) at level $k+1$ and construct a forest with $i_{k} - i_{k-1}$ trees, having a total of $i_{k}$ internal nodes. Note that $(2i_{k}-i_{k+1})-(2i_{k+1}-i_{k+2}) \geq 0$ and $(2i_{k}-i_{k+1}) \leq n$, hence, we have sufficient characters to create such a forest. This proves the theorem.
\end{proof}
Theorem \ref{valid} follows when $k=0$ and $i_0=n-1$. Note that the above conditions are necessary for any tree as the number of characters below a level $l$($2i_l-i_{l+1}$) in a tree decreases and the number of internal nodes at or below a level $l$($i_l$) in a tree decreases with increasing $l$.
}

\subsection{\label{appendix_exact_genllhc_main}Exact DP for \textbf{\GSLLHC}: Proof of Theorem~\ref{thm:exact} {\em \bf (a)}}
\label{sec:proof3a_main}

We start with a simple proposition.
\begin{proposition}
\label{optimality_main}
A character having higher frequency will appear at the same or lower level
(that is  closer to the root) than a character having lower frequency.
\end{proposition}
The proposition is easy to verify - if this was not true, one could simply swap
the two characters thereby improving the objective value as well as the penalty.

Note that for the {\GSLLHC} problem also,
the code length can be represented as sum of $h$ prefix-sums.
We will now show that for {\GSLLHC}(${\cal P}, \pnlty(\cdot), f(\cdot)$), the objective value $\objT(T)$, for any tree $T$, can also be written as a sum of $h$ terms, 
where each term corresponds to the contribution by the corresponding level, to the objective value, of the tree. Recall that $f(\cdot)$ was a monotonically non-decreasing function.
This result is captured in Lemma~\ref{lem:dectime_main}.
We define two new function $\hat{f}(.),\hat{p}(.)$:
$$\hat{f}(i) =
\left\{
	\begin{array}{ll}
		f(1)  & \mbox{if } i = 1 \\
		f(i)-f(i-1) & \mbox{if } i > 1
	\end{array}
\right.
$$
$$\hat{p}(i) =
\left\{
	\begin{array}{ll}
		p(1)  & \mbox{if } i = 1 \\
		p(i)-p(i-1) & \mbox{if } i > 1
	\end{array}
\right.
$$

Now if $h$ represents the total height of a tree, $T$, then we have the following lemma. 
\begin{lemma}
\label{lem:dectime_main}
\end{lemma}
$$\objT(T) = \sum_{\lvl=0}^{h-1} \hat{f}({\lvl+1})\cdot (\prefx_{2 i_\lvl - i_{\lvl+1}})$$

\begin{proof}
    $$\objT(T) = \sum_{\lvl=0}^{h-1} \hat{f}({\lvl+1})\cdot (\prefx_{2 i_\lvl - i_{\lvl+1}})$$

By definition of ${\objT(T)}$ we have
\begin{eqnarray*}
\objT(T) & = & \sum_{c \in \alphabet} \left( \freq(c) \cdot \objc(d_T(c)) \right) \\
\end{eqnarray*}
By using the definition of $\hat{f}(i)$ we get
$$\objT(T)=\sum_{c \in \alphabet} \left( \freq(c) \cdot \left(\sum_{i \le \dT(c)} \hat{f} (i) \right) \right)$$
By rearranging the summation over each level and using proposition \ref{charbelow} we get
$$\objT(T)=\sum_{\lvl=0}^{h-1} \left( \hat{f} (\lvl+1) \cdot \sum_{j=1}^{2 i_\lvl - i_{\lvl+1}}freq(j) \right)$$
and using Theorem~\ref{prefixtree} we get
$$\objT(T)=\sum_{\lvl=0}^{h-1} \hat{f} ({\lvl+1}) \cdot \prefx_{2 i_\lvl - i_{\lvl+1}}$$

This completes the proof of the Lemma.
\end{proof}


Using Lemma \ref{lem:dectime_main}, it remains to find a sequence of $i_l$'s as before except that now the goal is to minimize the
objective value, $\objT(T)$, of the tree. 
The prefix tree can be constructed by alluding to Theorem~\ref{valid}. We describe a recurrence to obtain such a sequence.
Let $D(i,\lvl,P)$ denote the minimum objective value amongst all forests rooted at level $\lvl$,
having $i$ internal nodes with penalty at most $P$ 
(here, the objective value of the forest is the sum of the objective values of the trees in the forest).

A dynamic program using the above recurrence can be designed as follows. 
Let $h$ be some upper bound on the height of the optimal prefix tree.

{\em Base Case:}
For all forests with no internal nodes, we initialize the objective value to $0$, i.e.,
$$ \forall ~\lvl \in [0,h] ~and ~P\in[0,{\cal P}]:~~D(0,\lvl,P)=0$$

{\em Inductive Case:}
To compute $D(i,\lvl,P)$, we iterate over the number of internal nodes at depths greater than $\lvl$.
If $j$ internal nodes are at depths strictly greater than $\lvl$, 
then there are $(2i-j)$ characters at depths strictly greater than $\lvl$ and 
$\hat{f}({\lvl+1}) \cdot \prefx_{2i-j}$ is the contribution, to the objective value $\objT(T)$,
of all the characters having level(depth) $>\lvl$, due to the access at level $(\lvl+1)$.
Furthermore, $D(j,\lvl+1,P')$ denotes the objective value contributed by all accesses made at levels(depths) greater than $\lvl+1$. 
This yields the following recurrence:
\begin{eqnarray}
\label{eq:rec0_main}
D(i,\lvl,P) = \min_{\substack{ j \in [max(0,2i-n),i-1] \\ \& \ 2i-j \ \ge \ 2j-k}} \left\{ D(j,\lvl+1,P')  + \hat{f}({\lvl+1})\cdot \prefx_{2 i-j} \right\} & &
\end{eqnarray}
where $P' = P-\hat\pnlty(\lvl+1)\cdot\prefx_{2i-j}$ and $k$ is the recursive index using which $D(j,\lvl+1,P')$ was populated. We only need to recurse if $P'>0$. The tree with the optimal objective value can be obtained by maintaining the parent pointers of each update and backtracking. We present the pseudo-code in Algorithm \ref{alg:exact}.
\begin{algorithm}[t!]
\DontPrintSemicolon 
\KwIn{Weighted Alphabet $\alphabet=\{c_1, c_2, \ldots, c_n\}$; Penalty bound ${\cal P}$; penalty function \pnlty(.); objective function $f(.)$; function $\hat{f}(.)$ defined over $f(.)$}
\KwOut{Minimum objective value among trees having penalty less than ${\cal P}$}
\For{$\lvl \gets 0 ~\textbf{to}~ h$ }{
    \For{$P \gets 0 ~\textbf{to}~ {\cal P}$ }{
        $D(0,\lvl,P):=0$        
    }
}
\For{$i \gets ~1~\textbf{to}~(n-1)$}{
    \For{$\lvl \gets (h-1) ~\textbf{downto}~ 0$ }{
        \For{$P \gets 0 ~\textbf{to}~ {\cal P}$ }{
            $bestObjVal:=\infty $;\\
            \For{$j \gets \max(0,2i-n) ~\textbf{to} ~i-1$}{    
                $objVal:= \infty $\\
                $P' = P - \pnlty(\lvl+1)\cdot \prefx_{2i-j}$\\
                $k$ $\gets$ recursive index where $D(j,\lvl+1,P')$ was minimized\\
                \uIf{$2i-j < 2j-k$}{
                    continue;
                }
                \uIf{$P' > 0$}{$objVal:=D(j,\lvl+1,P')+\hat{f}({(l+1)})\cdot \prefx_{2 i-j}$}
                
                \uIf{$objVal < bestObjVal $}{
                    $bestObjVal:=objVal$
                }
            }
            $D(i,\lvl,P):=bestObjVal$
        }
    }
}
\Return{$D(n-1,0,{\cal P})$};
\caption{(for \textbf{Theorem \ref{thm:ptas}(a)})}
\label{algo:doptDP}
\label{alg:exact}
\end{algorithm}


{\em Time complexity: }
There are  $O(n)$ characters, $h$ levels and $O({\cal P})$ values for penalty;
hence there are $O(nh {\cal P})$ cells in the table. As each cell can be filled in $O(n)$ time, the time complexity is $O(n^2 h {\cal P})$.
As height, $h$ is at most $n$, we get the time complexity to be $O(n^3 {\cal P})$.
As ${\cal P}$ may not be polynomial in $n$, this is a pseudo-polynomial time algorithm. 

As the above algorithm is symmetric in terms of penalty and objective value, we can find the tree having the minimum penalty and objective value at most ${\cal C}$ in  $O(n^3 {\cal C})$ time using the recurrence 
\begin{eqnarray}
\label{eq:rec10_main}
D(i,\lvl,C) = \min_{\substack{ j \in [max(0,2i-n),i-1] \\ \& \ 2i-j \ \ge \ 2j-k}} \left\{ D(j,\lvl+1,C-\prefx_{2i-j}) + \hat{p}({\lvl+1})\cdot \prefx_{2 i-j} \right\} & &
\end{eqnarray}
Let the penalty of the solution to the above $DP$ be ${\cal{P}}_{dual}$, we will use it to give a $PTAS$ algorithm for $\GSLLHC$ in the next section.

\eat{
\begin{algorithm}[t!]
\DontPrintSemicolon 
\KwIn{Weighted Alphabet $\alphabet=\{c_1, c_2, \ldots, c_n\}$; Penalty bound ${\cal P}$; penalty function \pnlty(.); objective function $f(.)$; function $\hat{f}(.)$ defined over $f(.)$}
\KwOut{Minimum objective value among trees having penalty less than ${\cal P}$}
\For{$\lvl \gets 0 ~\textbf{to}~ h$ }{
    \For{$P \gets 0 ~\textbf{to}~ {\cal P}$ }{
        $D(0,\lvl,P):=0$        
    }
}
\For{$i \gets ~1~\textbf{to}~(n-1)$}{
    \For{$\lvl \gets (h-1) ~\textbf{downto}~ 0$ }{
        \For{$P \gets 0 ~\textbf{to}~ {\cal P}$ }{
            $bestObjVal:=\infty $;\\
            \For{$j \gets \max(0,2i-n) ~\textbf{to} ~i-1$}{    
                $objVal:= \infty $\\
                $P' = P - \pnlty(\lvl+1)\cdot \prefx_{2i-j}$\\
                $k$ $\gets$ recursive index where $D(j,\lvl+1,P')$ was minimized\\
                \uIf{$2i-j < 2j-k$}{
                    continue;
                }
                \uIf{$P' > 0$}{$objVal:=D(j,\lvl+1,P')+\hat{f}({(l+1)})\cdot \prefx_{2 i-j}$}
                
                \uIf{$objVal < bestObjVal $}{
                    $bestObjVal:=objVal$
                }
            }
            $D(i,\lvl,P):=bestObjVal$
        }
    }
}
\Return{$D(n-1,0,{\cal P})$};
\caption{(for \textbf{Theorem \ref{thm:ptas}(a)})}
\label{algo:doptDP}
\label{alg:exact}
\end{algorithm}
}

\subsection{\label{appendix_ptas_genllhc_main}PTAS for \textbf{\GSLLHC}: Proof of Theorem~\ref{thm:ptas} {\em \bf (b)} and  Theorem~\ref{gllhc2}}
\label{sec:proof3b_main}
We first prove Theorem \ref{gllhc2}. Theorem~\ref{thm:ptas}(b) follows as $h$ takes value at most $n$.

The algorithm presented in the previous section has linear running time dependency on the parameter ${\cal P}$. In this section, we propose a polynomial time approximation algorithm that runs in time $O(n^4/\epsilon)$
and returns a prefix tree having penalty at most ${\cal P}$ and objective value with in $(1+\epsilon)$ times the optimal value. We first give an algorithm which returns a tree with penalty at most the value of the minimum penalty possible for tree with objective value at most ${\cal C}$, and objective value at most $(1+\epsilon){\cal C}$.

For this we restrict the parameter $C$ to only take on values that are multiples of $\lambda = \lfloor (\epsilon \cdot {\cal C})/2h \rfloor$
ranging from $0 \cdot \lambda$ upto $((2h/\epsilon) + h) \cdot \lambda$ 
where $h$ is some upper bound on the height of the optimal prefix tree.
We denote the dynamic program table maintained by this algorithm with $\overline{D}$. Let $\overline{D}(i,\lvl,C)$ denote the minimum penalty amongst all forests rooted at level $\lvl$,
having $i$ internal nodes with objective value at most $C$ 
(here, the penalty of the forest is the sum of the penalties of the trees in the forest). Note, here each $DP$ cell stores a structure having minimum penalty as compared to {the exact algorithm of \GSLLHC}, where each $DP$ cell stores a structure having minimum objective value.
 
We define a rounding function ${\mathbf r}$ as follows:
$${\mathbf r}(x) = \left\lceil \dfrac{x}{\lambda} \right\rceil \cdot \lambda.$$
We change the recurrence from the previous section as follows:
The base case becomes:
for all forests with no internal nodes, we initialize the objective value to $0$, i.e.,
$\forall ~\lvl \in [0,h]$ and $C$ a multiple of $\lambda$ and $C$ $\in[0,{\mathbf r}({\cal C})+h\lambda]$:
$D(0,\lvl,C)=0.$. The inductive step is modified to:
\begin{eqnarray}
\label{eq:rec_main}
\overline{D}(i,\lvl,C) = \min_{\substack{ j \in [max(0,2i-n),i-1] \\ \& \ 2i-j \ \ge \ 2j-k}} \left\{ \overline{D}(j,\lvl+1,C')  + \hat{p}_{\lvl+1}\cdot \prefx_{2 i-j} \right\} & &
\end{eqnarray}
where $C' = C-{\mathbf r}(\hat{f}_{\lvl+1}\cdot \prefx_{2 i-j})$ and $k$ is the recursive index using which $\overline{D}(j,\lvl+1,C')$ was populated.
Note that we only update entries of $\overline{D}$ for which the $C$ parameter is itself a multiple of $\lambda$.
It is easy to see that the $C$ parameter will take on only $O(h/\epsilon)$ values. 
The table can be compressed accordingly and maintained only for these entries,
however we omit these implementation details in the interest of better readability. We present the Pseudo-code for \label{appendix_ptas_genllhc}PTAS for \textbf{\GSLLHC} in Algorithm \ref{alg:pseudo}.
\begin{algorithm}[ht!]
\DontPrintSemicolon 
\KwIn{Weighted Alphabet $\alphabet=\{c_1, c_2, \ldots, c_n\}$; Penalty bound ${\cal P}$; penalty function \pnlty(.); objective function $f(.)$; function $\hat{f}(.)$ defined over $f(.)$; Approximation constant $\epsilon$}
\KwOut{Prefix tree having penalty $ \leq {\cal P}$ and objective value less than $\leq (1+\epsilon)\cdot C^*$}
$C_{PTAS}$ $\gets$ $\infty$\\

\For{$val \gets 0 ~\textbf{to}~ \log_{2}{({\cal F}\cdot f(n))}$}{
$\cal{C}$ $\gets$ $2^{val}$\\
$\lambda = \lfloor (\epsilon \cdot {\cal C})/2h \rfloor$\\
\For{$\lvl \gets 0 ~\textbf{to}~ n$ }{
    \For{$b \gets 0 ~\textbf{to}~ ((2h/\epsilon) + h)$}{
        $C=b\cdot \lambda$\\
        $\overline{D}(0,\lvl,C):=0$    
    }
}
\For{$i \gets ~1~\textbf{to}~(n-1)$}{
    \For{$\lvl \gets (h-1) ~\textbf{downto}~ 0$ }{
        \For{$b \gets 0 ~\textbf{to}~ ((2h/\epsilon) + h)$ }{
            $C=b\cdot \lambda$\\
            $bestPenalty:=\infty $\\
            \For{$j \gets \max(0,2 i-n) ~\textbf{to}~ i-1$}{    
                $Penalty:= \infty $\\
                $C' = C - {\mathbf r}(\pnlty(\lvl+1)\cdot \prefx_{2 i-j})$\\
                $k$ := recursive index where $\overline{D}(j,\lvl+1,C')$ was minimized\\
                \uIf{$2i-j < 2j-k$}{
                    continue;
                }
                \uIf{$0\leq C'$}{$Penalty:=\overline{D}(j,\lvl+1,C')+\hat{f}_{(l+1)}\cdot \prefx_{2i-j}$}
                
                \uIf{Penalty $<$ bestPenalty and Penalty $< \cal{P}$}{
                    $bestPenalty:=Penalty$
                }
            }
            $\overline{D}(i,\lvl,C):=bestPenalty$
        }
    }
}

$C_{PTAS}:=\min_{C}(\overline{D}(n-1,0,C)~corresponds~to~a~valid~prefix~tree)$\\
\uIf{$C_{PTAS} \leq \cal{C}$}{
    \textbf{break}
}
}
\Return{$C_{PTAS}$};
\caption{(for {\bf Theorem \ref{thm:ptas}(b)})}
\label{algo:doptptasDP}
\label{alg:pseudo}
\end{algorithm}

\eat{
\begin{algorithm}[t!]
\DontPrintSemicolon 
\KwIn{Weighted Alphabet $\alphabet=\{c_1, c_2, \ldots, c_n\}$; Penalty bound ${\cal P}$; penalty function \pnlty(.); objective function $f(.)$; function $\hat{f}(.)$ defined over $f(.)$; Approximation constant $\epsilon$}
\KwOut{Prefix tree having penalty $ \leq {\cal P}$ and objective value less than $\leq (1+\epsilon)\cdot C^*$}
$C_{PTAS}$ $\gets$ $\infty$\\

\For{$val \gets 0 ~\textbf{to}~ \log_{2}{({\cal F}\cdot f(n))}$}{
$\cal{C}$ $\gets$ $2^{val}$\\
$\lambda = \lfloor (\epsilon \cdot {\cal C})/2h \rfloor$\\
\For{$\lvl \gets 0 ~\textbf{to}~ n$ }{
    \For{$b \gets 0 ~\textbf{to}~ ((2h/\epsilon) + h)$}{
        $C=b\cdot \lambda$\\
        $\overline{D}(0,\lvl,C):=0$    
    }
}
\For{$i \gets ~1~\textbf{to}~(n-1)$}{
    \For{$\lvl \gets (h-1) ~\textbf{downto}~ 0$ }{
        \For{$b \gets 0 ~\textbf{to}~ ((2h/\epsilon) + h)$ }{
            $C=b\cdot \lambda$\\
            $bestPenalty:=\infty $\\
            \For{$j \gets \max(0,2 i-n) ~\textbf{to}~ i-1$}{    
                $Penalty:= \infty $\\
                $C' = C - {\mathbf r}(\pnlty(\lvl+1)\cdot \prefx_{2 i-j})$\\
                $k$ := recursive index where $\overline{D}(j,\lvl+1,C')$ was minimized\\
                \uIf{$2i-j < 2j-k$}{
                    continue;
                }
                \uIf{$0\leq C'$}{$Penalty:=\overline{D}(j,\lvl+1,C')+\hat{f}_{(l+1)}\cdot \prefx_{2i-j}$}
                
                \uIf{Penalty $<$ bestPenalty and Penalty $< \cal{P}$}{
                    $bestPenalty:=Penalty$
                }
            }
            $\overline{D}(i,\lvl,C):=bestPenalty$
        }
    }
}

$C_{PTAS}:=\min_{C}(\overline{D}(n-1,0,C)~corresponds~to~a~valid~prefix~tree)$\\
\uIf{$C_{PTAS} \leq \cal{C}$}{
    \textbf{break}
}
}
\Return{$C_{PTAS}$};
\caption{(for {\bf Theorem \ref{thm:ptas}(b)})}
\label{algo:doptptasDP}
\label{alg:pseudo}
\end{algorithm}
}

The following Lemma shows we can get a prefix-tree with penalty $\le$ ${\cal{P}}_{dual}$ by sacrificing 
an additive $\lambda$ factor for every level in the objective value.
\begin{lemma}\label{lem:ptas_ineq}
For any valid values of $i$, $\lvl$ and $P$: \ \ 
$D(i,\lvl, C) \ \ge \ \overline{D}(i,\lvl,{\mathbf r}(C)+(h-\lvl)\cdot \lambda)$
\end{lemma}
\begin{proof}
For any valid values of $i$, $\lvl$ and $P$:
$$D(i,\lvl, C) \ \ge \ \overline{D}(i,\lvl,{\mathbf r}(C)+(h-\lvl)\cdot \lambda)$$

In our proof we will be using the following fact:
\begin{proposition}
\label{prop:req1}
Let $C'$ and $C''$ be multiples of $\lambda$. Then,
$\overline{D}(z,\lvl+1,C') \ge \overline{D}(z,\lvl+1,C'')$ whenever $C' \le C''$.
\end{proposition}
This proposition holds because the best solution having objective value at most $C'$ is also a candidate solution
having objective value at most $C''$ (other parameters remaining same).

We now prove the lemma by induction on the value of $\lvl$ decreasing from $h$ to $0$.

\par\noindent
For $\lvl=h$: 
From our initialization, the entries of $D$ and $\overline{D}$
are all initialized to $0$ for $\lvl=h$ and hence the claim trivially holds.

\par\noindent
For $\lvl < h$:
Consider $D(i,\lvl,C)$.
From recurrence~(\ref{eq:rec10_main}), there must be some choice of $j$ for which
$D(i,\lvl,C)$ is minimized. Let $z$ be that choice of $j$, i.e.,
$$D(i,\lvl,C) = D(z,\lvl+1,C-\hat{f}_{\lvl+1}\cdot \prefx_{2i-z}) + \hat{p}_{\lvl+1} \cdot \prefx_{2i-z} $$

Now we obtain the following relations:
\begin{flalign*}
& D(i,\lvl,C) & \\
& \ \ = D(z,\lvl+1,C-\hat{f}_{\lvl+1}\cdot\prefx_{2i-z}) + \hat{p}_{\lvl+1} \cdot \prefx_{2i-z} & \\
& \ \ \ge \overline{D}(z,\lvl+1,{\mathbf r}(C-\hat{f}_{\lvl+1}\cdot \prefx_{2i-z})+(h-(\lvl+1))\lambda) + \hat{p}_{\lvl+1} \cdot \prefx_{2i-z} & \\
& \ \ \ge \overline{D}(z,\lvl+1,{\mathbf r}(C)-({\mathbf r}(\hat{f}_{\lvl+1}\cdot \prefx_{2i-z})-\lambda)+(h-(\lvl+1))\lambda) + \hat{p}_{\lvl+1} \cdot \prefx_{2i-z} & \\
& \ \ = \overline{D}(z,\lvl+1,{\mathbf r}(C)-{\mathbf r}(\hat{f}_{\lvl+1}\cdot \prefx_{2i-z})+(h-\lvl)\lambda) + \hat{p}_{\lvl+1} \cdot \prefx_{2i-z} & \\
& \ \ \ge \overline{D}(i,\lvl,{\mathbf r}(C)+(h-\lvl)\lambda) & 
\end{flalign*}
where the first inequality follows by induction,
the second inequality follows from Proposition~\ref{prop:req1}
and the last inequality follows from the fact that
$\overline{D}(z,\lvl+1,{\mathbf r}(C)-{\mathbf r}(\hat{f}_{\lvl+1}\cdot \prefx_{2i-z})+(h-\lvl)\lambda) +  \hat{p}_{\lvl+1} \cdot \prefx_{2i-z} $ is also a candidate for consideration in recurrence~(\ref{eq:rec_main}) for $\overline{D}(i,\lvl,{\mathbf r}(C)+(h-\lvl)\lambda)$.

This completes the proof of the Lemma.
\end{proof}
From the previous section, we know that the optimal solution is captured by $D(h-1,0,{\cal C})$.
Hence Lemma \ref{lem:ptas_ineq} implies that the optimal solution is also captured by $\overline{D}(h-1,0,{\mathbf r}({\cal C}) + h \cdot \lambda )$.
We now look at all the entries $\overline{D}(h-1,0,{\mathbf r}({ C}) + h \cdot \lambda ) \leq P_{dual}$ and pick the entry with the minimum value of ${\mathbf r}({C}) + h \cdot \lambda $. 

Hence using $\overline{D}$, given a objective function threshold $\cal{C}$, we can find a prefix tree with penalty  $\leq P_{dual} \leq {\cal P}$ and objective value $\leq (1+\epsilon)\cdot{\cal{C}}$. Now, if substitute $\cal{C}$ $=C^{*}$, where $C^{*}$ is the objective value of the solution to the {\GSLLHC}  problem, we will have the solution to the $PTAS$ of {\GSLLHC} problem. Now, instead of calculating $C^{*}$ directly we use binary search in the range $[0, {\cal{F}}\cdot f(n)]$, where $\cal{F}$ is the cumulative frequency of all the characters in the prefix tree and $f(n)$ is the value of objective function at level $n$. As the depth is at most $n$ for all characters and $f(.)$ is an increasing function, the objective value is at most ${\cal{F}}\cdot f(n)$. Thus, we have a $PTAS$ algorithm for \GSLLHC.

{\em Time complexity:} 
There are at most $O(h)$ levels and $O(n)$ characters. $C$ can have at most $O(h/ \epsilon)$ possible values. 
Hence, there are $O(nh^2/\epsilon)$ cells in the table. 
Each cell can be filled in at most $O(n)$ time. 
So the time complexity is $O({n^2h^2}/{\epsilon})$. 
Since there are a total of $h$ recursive calls, the error in objective function value is bounded by $h\lambda \le \epsilon \cdot {\cal C}$.
Thus, we can find a prefix tree having objective value less than $(1+\epsilon)\cdot C^*$ and penalty at most ${\cal P}$ in $O( {n^2h^2}/{\epsilon} )$ time. This proves Theorem \ref{gllhc2}. Taking the upper bound for the height, $h$, as $n$, we get the time complexity to be $O( {n^4}/{\epsilon} )$, which proves Theorem \ref{thm:ptas}(b).

\section{Algorithms for the Code Optimal Prefix Tree (\DOPT) problem}
\label{copt_algorithms}
For a fixed number of block levels, $m$, the possible number of values corresponding to 
the decode time for the forests in the dynamic program is $n^{m-1}$ \eat{(similar to Algorithm \ref{alg:exact})}. 
We use this to give an $O(n^{m+2})$ algorithm for Theorem \ref{thm:constanthei}(a).
\subsection{Proof of Theorem \ref{thm:memhei} (a) }
There exists a dynamic program algorithm to solve the {\DOPT($\cal P$)} problem that runs in time
$O(n^{2+m})$ for $m$ block levels.

The dynamic programming algorithm\eat{(Algorithm~\ref{algo:doptdualDP})} is similar to that for the exact algorithm \eat{(Algorithm~\ref{algo:doptDP})}
with the main difference being that instead of iterating over lengths we iterate over the decode times of the tree.

Let $\dual(i,\lvl,T)$ denote the minimum codelength amongst all forests rooted at level $\lvl$,
having $i$ internal nodes with decode time at most $T$. Also, define ${\cal T}=\sum\limits_{c=1}^{n}\cdot  \sum\limits_{i=1}^{m}q_i$ 
(here, the decode time of the forest is the sum of the decode times of the trees in the forest)

For a fixed number of block levels, $m$, the following lemma holds:
\begin{lemma}\label{seq_of_xi}
The number of possible values of decode time for the forests rooted at some level is  $n^{m-1}$.
\end{lemma}
\begin{proof}
Let there be $x_i$ characters in the $i$th block level $\forall i\in[1,m]$ and $x_0$ be the number of characters which are not present in the forest corresponding to $\dual(i,\lvl,T)$. These characters corresponding to $x_0$ will not have any decode time contribution for $T$. We have $\sum_{i=0}^{m} x_i = n$. For $l>w_1$, the width of first block, we know that $x_1$ is zero. When $l\leq w_1$, we know that $x_0$ is zero. That is not both of $x_0,x_1$ can be non-zero. Hence, there are $O(n^{m-1})$ possible sequences of $x_i$'s satisfying this. For each sequence of $x_i$'s, we can uniquely determine the decode time value. Hence there are $O(n^{m-1})$ possible decode time values.
\end{proof}

A dynamic program using the above recurrence can be designed as follows:

{\em Base Case:}
For all forests with no merges, i.e. no internal nodes, we initialize the decode time to $0$, i.e.,
$$ \forall ~\lvl \in [0,n] ~and ~T\in[0,{\cal T}]:~~\dual(0,\lvl,T)=0$$

{\em Inductive Case:}
To compute $\dual(i,\lvl,T)$, we iterate over the number of internal nodes that are at depth strictly greater than $\lvl$.
If $j$ internal nodes are at depth strictly greater than $\lvl$, then there are $(2i-j)$ characters at depth strictly greater than $\lvl$, then $q_{\lvl+1}\cdot P_{2i-j}$ is the decode-time contribution
of all the characters having level $>\lvl$, due to the access at level $(\lvl+1)$. 
Furthermore, $\dual(j,\lvl+1,T')$ denotes the decode time contributed by all accesses made at depths greater than $\lvl+1$. 
This yields the following recurrence:
\begin{eqnarray}
\label{eq:recA}
\dual(i,\lvl,T) =  \min_{\substack{ j \in [max(0,2i-n),i-1] \\ \& \ 2i-j \ \ge \ 2j-k}} \left\{ \dual(j,\lvl+1,T')  +  P_{2\cdot i-j} \right\} & &
\end{eqnarray}
where $T' = T - \hat{q}_{lvl+1}\cdot P_{2\cdot i-j} $ and $k$ is the recursive index using which $\dual(j,\lvl+1,T')$ was populated. We only need to recursively check if $T'>0$. The tree with the optimal decode time can be obtained by maintaining the parent pointers of each update and then backtracking.

Note that we only update entries of $\dual$ for which the $T$ parameter corresponds to a sequence of  $\langle$ $x_0,x_1,x_2,\ldots,x_m$ $\rangle$, from lemma \ref{seq_of_xi}.
The decode time for a sequence $\langle x_0,x_1,x_2,\ldots,x_m \rangle$ is $Dec(\langle x_i \rangle)= \sum_{i=1}^{m} x_i\cdot \hat{q}_i$

From Lemma \ref{seq_of_xi}, we know that the $T$ parameter will take on only $O(n^{m-1})$ values. 
The table can accordingly be compressed and maintained only for these entries,
however we omit these implementation details in the interest of better exposition.

After the DP is filled, we check all the entries of the form $\dual(n-1,0,t)$ which have code length parameter $t \leq {\cal P}$ and find the the optimal code length corresponding to it.

{\em Time complexity}: We note that $i$ and $\ell$ can take $n$ possible values each and $T$ takes $n^{m-1}$ possible values.
So, there are $n^{m+1}$ cells in the DP. Each cell can be filled in at most $O(n)$ time. So, the time complexity of the DP is $O(n^{m+2})$. Checking the DP table to find the optimal decode time will take $O(n^{m+1})$ time. 
Hence, we can solve the $COPT$ problem in $O(n^{m+2})$ time when the number of block levels is a constant $m$.

\begin{algorithm}[t!]
\DontPrintSemicolon 
\KwIn{Alphabet $\alphabet=\{c_1, c_2, \ldots, c_n\}$; with $frequencies=\{f_1, f_2, \ldots, f_n\}$; Decode time bound ${\cal P}$}
\KwOut{Minimum possible code length}
\For{ $\lvl \gets 0 ~\textbf{to}~ n$}{
    \For{\textbf{all} sequence $<x_i>$ satisfying Lemma~\ref{seq_of_xi}}{
        $T \gets Dec(<x_i>) $\\
        $\dual(0,\lvl,T):=0$        
    }
}
\For{$i \gets ~1~\textbf{to}~(n-1)$}{
    \For{$\lvl \gets (n-1) ~\textbf{downto}~ 0$ }{
        \For{\textbf{all} sequence $<x_i>$ satisfying Lemma~\ref{seq_of_xi}}{
            $bestLen:=\infty $;\\
            $T \gets Dec(<x_i>) $\\
            \For{$j \gets \max(0,2 i-n) ~\textbf{to}~i-1$}{    
                $codLen:= \infty $\\
                $T' = T - \hat{q}_{lvl+1}\cdot P_{2i-j}$\\
                $k$ := recursive index where $\dual(j,\lvl+1,T')$ was minimized\\
                \uIf{$2i-j < 2j-k$}{
                    continue;
                }
                \uIf{$0\leq T'$}{$codLen:=\dual(j,\lvl+1,T')+ P_{2 i-j}$}
                
                \uIf{$codLen < bestLen $}{
                    $bestLen:=codLen$
                }
            }
            $\dual(i,\lvl,T):=bestLen$
        }
    }
}
$minCodeLength=\min (\infty, \min_{t : t \leq {\cal P}}(\dual(n-1,0,t)))$\\
\uIf{$minCodeLength = \infty$}{
    \Return{"No Valid Tree"}
}
\Return{$minCodeLength$};
\caption{(for \textbf{Theorem \ref{thm:memhei}(a)})}
\label{algo:doptdualDP}
\end{algorithm}

We now present the proof of Theorem~\ref{thm:memhei}(b).
\subsection{Proof of Theorem \ref{thm:memhei}(b)}
\label{sec:ptas_for_memhei}
Consider the blocking scheme in the definition of the $\DOPT$ problem. 
As mentioned in the introduction,
the number of block levels, $m$,
is typically a small constant in practice. 
We now present a more efficient dynamic program based pseudo-approximation algorithm for the case 
when the number of block levels is constant.

We first prove some results (c.f. Propositions~\ref{prop:comp}, \ref{prop:squash} and Lemma~\ref{lem:dist}) 
required in the formulation of our new dynamic program.
The following proposition shows that given a set of characters,
we can construct a (nearly complete) prefix tree of depth $\lceil \log n \rceil$.
\begin{proposition}
\label{prop:comp}
There exists a prefix tree for a set of characters, $C$, having depth $\lceil \log |C| \rceil$.
\end{proposition}
\begin{proof}
It is easy to verify that we can place $2^{\lceil \log |C| \rceil}-|C|$ characters at depth 
$\lceil \log |C| \rceil - 1$ in the subtree and the remaining characters
at depth $\lceil \log |C| \rceil$ to form a valid prefix tree 
(additional nodes are added to serve as internal nodes).
\end{proof}
Consider a set of characters, $C$. 
The following proposition shows that given an arbitrary tree, $T$, with characters of $C$ appearing as leaf nodes in $T$,
we can always construct a valid prefix tree over $C$ that has height no more than that of $T$ and in which each character
appears at a depth no more than its depth in $T$.

\begin{proposition}
\label{prop:squash}
Let $C$ denote a set of characters. 
Given a tree, $T$, in which the characters of $C$ appear as leaf nodes,
there exists a valid prefix tree, $T'$, over $C$ that has no greater height than $T$ and 
in which $d_{T'}(c) \le d_T(c)$ $\forall c \in C$.
\end{proposition}
\begin{proof}
We start with the tree $T$ and iteratively modify it until we obtain a valid prefix tree.

We find a node(say $u$) that violates any of these conditions and modify the tree as follows:
\begin{itemize}
\item Is a leaf node but does not correspond to a character:
      We simply delete $u$.
\item Has only one child(say node $v$): 
      We remove $u$ and directly attach $v$ to the parent of $u$.
\end{itemize}
It is easy to see that when no more violating nodes are left, we get a valid prefix tree.
It is also straightforward to observe that we never increase the depth of any node in this process.
\end{proof}
The following Lemma shows that there cannot be too many levels in the optimal prefix tree 
between two consecutive characters of the alphabet when sorted in order of frequencies.
\begin{lemma}
\label{lem:dist}
{In a complete binary tree,} if $c_i$ is a character at level $\lvl$ and $c_{i+1}$ is at level $\lvl'$ then $\lvl'-\lvl$ $<$ $\lceil \log(n) \rceil$.
\end{lemma}
\begin{proof}
We prove this by contradiction. Let us assume that $\lvl'-\lvl$ {$\geq$} $\lceil \log(n)\rceil$. Since there is a leaf $c_{i+1}$ at depth greater than $\lvl$, there must be at least one internal node at the level $\lvl$. By our assumption there are no leaves in the tree rooted at this internal node, till the next $\lceil \log(n)\rceil$ levels. Hence there are at least $n$ internal nodes above level $\lvl'$. But the tree we started with has exactly $n-1$ internal nodes as it has $n$ leaves. Contradiction.
\end{proof}

The following lemma shows that there exists a tree having bounded height that has almost the same
code length and decode time as the optimal prefix tree of $\DOPT(\cal P)$.
\begin{lemma}
\label{lem-ht}
{Given $\delta > 0$}, there exists a prefix tree, $T'$, for which the code length is at most 
$(1+\delta)$ times the code length of $\DOPT(\cal P)$ and the height of $T'$ is no more than
$2m(\lceil {1}/{\delta} \rceil+ \lceil \log n \rceil)$ {, where $m$ is the number of block levels.}
\end{lemma}
\begin{proof}
Let $h^*$ denote the height (total number of tree levels) of the optimal prefix tree (solution to $\DOPT(\cal P)$). 
If $h^* \leq 2m(\lceil {1}/{\delta} \rceil+ \lceil \log n \rceil)$, then the
claim is trivially satisfied. We therefore focus on the case when
$h^* > 2m(\lceil {1}/{\delta} \rceil+ \lceil \log n \rceil)$.
As there are at most $m$ block levels, at least one of these has more than 
$2(\lceil {1}/{\delta} \rceil+ \lceil \log n \rceil)$ tree levels.

Let us focus on one such block level, {and let the starting tree level for the block level} be $\lvl'$. There must be at least one node $c_i$ between the tree levels $\lvl'+\lceil {1}/{\delta} \rceil+ \lceil \log n \rceil)$ and $\lvl'+\lceil {1}/{\delta} \rceil+ 2\lceil \log n \rceil)$ due to Lemma~\ref{lem:dist}. 
\begin{proposition}
\label{lem:dist2}
\label{lem:kth_highest_freq}
{In the optimal prefix tree, the $k$th highest frequency is at a level at most $k+\lceil \log(n) \rceil$.}
\end{proposition}
\begin{proof}
We prove this using induction. The base case holds from Lemma \ref{lem:dist}. Consider the two nodes at level one.

We first consider the case where not all $k$ highest frequencies are in the same sub-tree rooted at one of the nodes. By induction assumption, in the sub-tree in which $k$th highest frequency is present, the $k$th highest frequency is at level at most $k-1+\lceil \log(n) \rceil$. Therefore given holds.

We now consider the case where all $k$ highest frequencies are in the same sub-tree rooted at one of the nodes. Let the highest frequency in the other sub tree be $k+r'$th frequency for some $r'>0$. If $k+r'$th frequency is at level at most $k+\lceil \log(n) \rceil$, since higher frequencies are at a lower level, $k$th highest frequency is at level at most $k+\lceil \log(n) \rceil$. If not, the subtree has more than $2^{k+\lceil \log(n) \rceil -1}>n-1$ nodes. Contradiction.
\end{proof}

From the proposition, there are at least $\lceil {1}/{\delta} \rceil$ nodes with frequency higher than $c_i$.

Let $T^*$ be the prefix tree corresponding to the optimal solution $COPT({\cal P})$ and $\lvl = \lvl'+\lceil {1}/{\delta} \rceil+ \lceil \log n \rceil$.
We modify $T^*$ to construct another prefix tree, $T'$ as follows:
\begin{itemize}
\item all characters up to $\lvl+\lceil \log n \rceil$ retain the same level as in $T^*$, 
      except for $c_i$ 
\item $c_i$ is replaced with a new internal node, say $u$, and made a child of $u$ ($c_i$ is at level $\lvl+1$). 
\item We call a character of $T^*$ {\em deep} if it has depth more than
      $ 2m(\lceil {1}/{\delta} \rceil + \lceil \log(n) \rceil)$.
      Let $\gamma$ be the number of deep characters in $T^*$.
      Using Proposition~\ref{prop:comp}, there exists a subtree comprising of all the deep characters of $T^*$,
      having depth at most $\lceil \log \gamma \rceil \le \lceil \log n \rceil$.
      We attach this subtree as the second
      child of $u$. 
      The level of any of the characters in this subtree is no more than $\lvl+\lceil \log n \rceil + 1$.
\item We finally invoke Proposition \ref{prop:squash}, to get a valid prefix tree.
\end{itemize}

We now show that the codelength and decode time  of $T'$ are no more than $(1+\delta)$ times the corresponding parameters of $T^*$.
Note that both the code length and decode time of the deep characters of $T^*$ only reduces
as their depth reduces in $T'$. Therefore the code length and decode time can only increase due to the character $c_i$ moving 
one level (tree level) down. 

We first analyze the increase in code length due to $c_i$ moving one (tree) level down.
Recall that the block level to which $c_i$ belonged was divided into $2\lceil 1/\delta \rceil$ partitions
and $c_i$ belongs to the $\lceil 1/\delta \rceil^{th}$ partition. Moreover each partition contains a character.
Also, there are at least $\lceil {1}/{\delta} \rceil$ nodes with frequency higher than $c_i$ 
and hence have tree level same or above that of $c_i$.
Thus $len(T^*) \ge \lceil 1/\delta \rceil \cdot f_i$. The increase in code length incurred by moving 
$c_i$ down one tree level is $f_i$. 
Thus $f_i \le \delta \cdot ( \lceil 1/\delta \rceil \cdot f_i ) \le \delta \cdot len(T^*)$.
Therefore $len(T') \le (1+\delta)\cdot len(T^*)$.

As $c_i$ lies between the tree levels  $\lvl'+(\lceil {1}/{\delta} \rceil+ \lceil \log n \rceil)$ and $\lvl'+(\lceil {1}/{\delta} \rceil+ 2\lceil \log n \rceil)$, the next tree level to $c_i$ must also be in the same block level. Therefore $\Delta(T') \leq \Delta(T^*) \le {\cal P}$.
\end{proof}

Given the above Lemma, the algorithm is quite straightforward - 
we simply invoke Theorem~\ref{gllhc2}
by bounding the $h$ parameter by $2m(\lceil \log(n) \rceil + \left\lceil{{1}/{\delta}}\right\rceil)$.

{\em Time Complexity:} The analysis is same as that of the algorithm for Theorem~\ref{gllhc2}; we know that the time taken is $O(h^2n^2/\epsilon)$.Taking the bound on the height $h$ to be $2m(\lceil \log(n) \rceil + \left\lceil{{1}/{\delta}}\right\rceil)$ and setting $\delta$ to be $\epsilon$, the running time becomes $O\left(\dfrac{n^2\cdot m^2}{\epsilon}\left(\log^2(n)+{\dfrac{1}{\epsilon^2}}\right)\right)$. For constant $m$, this yields a complexity of $O\left(\dfrac{n^2}{\epsilon}max\left(\dfrac{1}{\epsilon^2}, {\log^2(n)}\right)\right)$.

\section{\label{appendix_fujiwara}Algorithm for Max-GHT: Proof of Theorem \ref{thm:fujiwara}}
\subsection{Introduction}
The problems GHT(Generalized Huffman Tree) and Max-GHT(Max Generalized Huffman 
tree) were formulated by Fujiwara and Jacobs\cite{FujiwaraJ14}. We first state 
their problem definitions.
\begin{Definition}
{\bf GHT} Given $n$ arbitrary functions $f_1,f_2\cdots f_n$ corresponding to 
$n$ leaves, the objective of GHT is to determine a binary tree $T$ with these 
$n$ leaves, such that $\sum_{i=1}^{i=n} f_i(d_i)$ is minimized, where the 
$i$th leaf is at depth $d_i$ in $T$.
\end{Definition}

\begin{Definition}
{\bf Max-GHT} Given $n$ arbitrary functions $f_1,f_2\cdots f_n$ corresponding to $n$ leaves, the objective of Max-GHT is to determine a binary tree $T$ with these $n$ leaves, such that $\max_{i=1}^{i=n} f_i(d_i)$ is minimized, where the $i$th leaf is at depth $d_i$ in $T$.
\end{Definition}

Fujiwara et al proved that Max-GHT and GHT are NP-hard for general functions 
$f_1,f_2\cdots f_n$. However, they also proved that if each $f_i$ is 
non-decreasing, then Max-GHT can be solved in $O(n^2\log{n})$ time. The 
complexity of GHT was unresolved, if $f_i$ is non-decreasing.

However, there is an implicit assumption in their hardness proof. They assume 
that there 
exists a solution which is a full binary tree(all internal nodes have exactly 
two children) for both GHT and Max-GHT. While this has to be true when the 
functions are non-decreasing, it need not be true when the functions are 
arbitrary. Consider the following simple counter example - when there are two 
leaves and for $i=1,2$ we have function values $f_i(1)=1$ and $f_i(2)=0$. The 
optimal solution(with zero cost for both GHT and Max-GHT) will have both 
leaves at level $2$ and hence such tree cannot be full binary. This re-opens 
the problems they posed and we present a simple $O(n^2)$ algorithm to convert 
Max-GHT and GHT with general functions to problems where Max-GHT and GHT have 
non-decreasing functions. As a direct consequence of this, we have an 
$O(n^2\log{n})$ algorithm to solve Max-GHT with general functions. Due to this 
reduction, we conclude that if GHT with non-decreasing functions can be solved 
in polynomial time, then GHT with general functions can also be solved in 
polynomial time. We also note that there is a solution with full binary tree for 
both GHT and Max-GHT with non-decreasing functions.
\subsection{Reduction}
\begin{lemma}
The GHT and Max-GHT problem with $n$ arbitrary functions $f_1,f_2\cdots f_n$, can be reduced to a problem with $n$ non-decreasing functions $g_1,g_2\cdots g_n$ in $O(n^2)$ time, where $g_i(j) = \min_{l=j}^{n} f_i(l)$
\end{lemma}
\begin{proof}
We update $g_i's$ in a bottom to top manner with $l^{th}$ entry as 
$\min{(f_i(l),g_i(l+1))}$ for $l<n$ and $g_i(n)=f_i(n)$. Hence the total time 
taken is $O(n)$ per function and $O(n^2)$ in total. It's easy to see that 
these functions evaluate to $g_i(j) = \min_{l=j}^{n} f_i(l)$ using induction.

For correctness, the key property we use is that there is an optimal tree 
which is a solution to GHT/Max-GHT, such that for any pair of depths 
$( d_1,d_2)$, if $d_1 < d_2$ and $f_i( d_1)\geq f_i( d_2)$, then the 
$i$th leaf can not be at $d_1$ for any $i$. This is due to  a simple exchange 
argument as we switch a node from $d_1$ to $d_2$, the Kraft sum decreases 
(hence the tree is feasible) and cost will not increase.(Note that if the 
Kraft sum for a given set of depths is less than $1$, we can always construct 
a binary tree with those function values) Therefore the $i$th leaf can not be 
at level $l$ if $g_i(l) \neq f_i(l)$. Hence, the structure of the optimal 
solution remains unchanged by changing the values for such $l$.
\end{proof}
We note that as the tree need not be full binary, the maximum height need not 
be $n$ like in \cite{FujiwaraJ14}. The above algorithm's correctness remains 
valid even when maximum height exceeds $n$ and the run time would be $O(m)$, 
where $m$ is the input size(previously $n^2$).
\section{Conclusion and open problems}
\label{sec:concl}
Motivated by many practical challenges in implementing compression, we 
introduce and study a novel variation of finding optimal prefix trees 
where one is allowed to deviate from the optimal code length within a specified
bound. This allows us to capture more generalized decoding costs for which we 
develop a bi-criterion framework and present efficient algorithms. 
An important application of this framework is to a natural class 
of memory access cost functions that use blocking and to the best of our 
knowledge, this is the first work that lays the theoretical foundations 
and present
a family of algorithms with provable guarantees. 
An open problem is to proving NP-hardness for the 
{\GSLLHC} problem that could be quite challenging as exemplified by Theorem \ref{thm:fujiwara}.
Another interesting future direction is to study the empirical performance of our algorithms
with real world data sets on practical systems with hierarchical memory; we anticipate 
promising results, similar to those obtained for a closely related 
variant in the hierarchical memory setting where the goal is to minimize the decode time and the average code length 
is bound by a threshold parameter\cite{ourarxiv}.

\section*{Acknowledgement}
We are grateful to an anonymous reviewer for suggesting the use of Monge property and simplifying the proofs of Theorem \ref{thm:memhei} and Theorem \ref{softllhc} in a previous version of this manuscript. 

\bibliographystyle{alpha}
\bibliography{refs}

\end{document}